\documentclass[useAMS,usenatbib]{mn2e}

\RequirePackage{lineno}
\usepackage{graphicx}
\usepackage{dcolumn}
\usepackage{bm}
\usepackage{amsmath}
\usepackage{amssymb}
\usepackage{mathrsfs}
\usepackage[x11names]{xcolor}
\usepackage{natbib}
\usepackage{graphicx}
\usepackage{subfigure}
\usepackage{caption}
\usepackage{color}
\usepackage{epsfig}
\usepackage{multirow}
\allowdisplaybreaks[1]

\newcommand{\mdot}{\dot M}

\newcommand{\beq}{\begin{equation}}
\newcommand{\eeq}{\end{equation}}

\newcounter{ichi}
\newcounter{ni}
\newcounter{san}
\newcounter{yon}
\title[A burst in a bubble for HEGFs and afterglows]
{A burst in a wind bubble and the impact on baryonic ejecta: high-energy gamma-ray flashes and afterglows from fast radio bursts and pulsar-driven supernova remnants}
\author[K. Murase et al.]
{Kohta Murase$^{1}$, Kazumi Kashiyama$^{2}$, and Peter M\'esz\'aros$^{1}$
\\
$^{1}$Center for Particle and Gravitational Astrophysics; Department of Physics; Department of Astronomy \& Astrophysics, \\ \ The Pennsylvania State University, University Park, PA 16802, USA\\
$^{2}$Einstein Fellow -- Theoretical Astrophysics Center, Department of Astronomy, University of California, Berkeley, CA 94720, USA\\
}

\begin{document}

\date{Submitted to arXiv: 28 March 2016}
\pagerange{\pageref{firstpage}--\pageref{lastpage}} \pubyear{2016}
\maketitle
\label{firstpage}

\begin{abstract}
Tenuous wind bubbles, which are formed by the spin-down activity of central compact remnants, are relevant 
in some models of fast radio bursts (FRBs) and super-luminous supernovae.  
We study their high-energy signatures, focusing on the role of pair-enriched bubbles produced by young magnetars, 
rapidly-rotating neutron stars, and magnetized white dwarfs. 
(i) First, we study the nebular properties and the conditions allowing for escape of high-energy gamma-rays and 
radio waves, showing that their escape is possible for nebulae with ages of $\gtrsim10-100$~yr. 
In the rapidly-rotating neutron star scenario, we find that radio emission from the quasi-steady nebula itself 
may be bright enough to be detected especially at sub-mm frequencies, which is relevant as a 
possible counterpart of pulsar-driven SNe and FRBs. 
(ii) Second, we consider the fate of bursting emission in the nebulae. 
We suggest that an impulsive burst may lead to a highly relativistic flow, which would interact with the nebula.  
If the shocked nebula is still relativistic, pre-existing non-thermal particles in the nebula can be significantly boosted 
by the forward shock, leading to short-duration (maybe millisecond or longer) high-energy gamma-ray flashes.  
Possible dissipation at the reverse shock may also lead to gamma-ray emission. 
(iii) After such flares, interactions with the baryonic ejecta may lead to afterglow emission with a duration of days to weeks.  
In the magnetar scenario, this burst-in-bubble model leads to the expectation that nearby ($\lesssim10-100$~Mpc)
high-energy gamma-ray flashes may be detected by the High-Altitude Water Cherenkov Observatory and the 
Cherenkov Telescope Array, and the subsequent afterglow emission may be seen by radio telescopes such as the Very Large Array.  
(iv) Finally, we discuss several implications specific to FRBs, including constraints on the emission regions and 
limits on soft gamma-ray counterparts. 
\end{abstract}


\section{Introduction}
Fast radio bursts (FRBs) are a new class of transients discovered in the last decade~\citep{Lorimer+07,
Keane+12,Thornton+13,Burke-Spolaor&Bannister14,Spitler+14,Ravi+15,Petroff+15,Masui+15,Champion+15}.  
They are very short ($\sim 1-10$~ms) and bright ($\sim 0.1-1$~Jy) sporadic events observed in the
$\sim1$~GHz bands showing a large dispersion measure (DM), ${\rm DM}\sim500-1000 \ \rm pc \ cm^{-3}$.
If the dispersion originates mainly from the intergalactic propagation~\citep{Ioka03,Inoue+04}, 
the sources can be at cosmological distance up to $z \sim 1$~\citep[but see, e.g.,][]{Loeb+14,Maoz+15}, 
and a high brightness temperature requires a coherent emission mechanism~\citep[e.g.,][]{Lorimer+07,
2008ApJ...682.1443L,Thornton+13,Katz14}. 
Interestingly, despite being relatively common~\citep[$\sim 10 \ \%$ of core-collapse supernovae;]
[]{Thornton+13,Keane&Petroff15,Law+15}, the origin of FRBs is still unknown. 

Various possibilities for the progenitors of FRBs have been proposed~
\citep[see, e.g.,][and references therein]{Kulkarni+14}.
Among them, extragalactic young neutron stars (NSs) including strongly magnetized NSs (so-called magnetars) 
have been considered as promising candidates~\citep{Popov&Postnov10,Lyubarsky14,2014A&A...562A.137F,
Cordes&Wasserman16,Connor+16}.  
NSs are expected to form as compact remnants of core collapse supernova (SN) explosions, and a significant 
fraction of the magnetic energy and/or rotation energy can be extracted from the young NSs whose
magnetic fields have not decayed yet.  
Compact merger models are also often discussed, where either NS or white dwarf (WD) or black hole (BH) are
involved in the merger system~\citep{2013PASJ...65L..12T,2014ApJ...780L..21Z,Kashiyama+13}.
For NS-NS, NS-BH, and WD-WD mergers, baryonic ejecta with $\sim{10}^{-5}-{10}^{-2}~M_\odot$ are expected to 
move with a fast velocity of $\sim0.05-0.3$~c.  

In either case, one naturally expects the formation of a tenuous wind bubble embedded in the baryonic ejecta, 
and the relativistic wind is driven by the spin-down activity of the NSs or WDs.  
It has been thought that pulsar winds that are initially Poynting-dominated become relativistic by the time 
they reach the light cylinder, and may further be accelerated in the wind zone via magnetic dissipation 
processes~\citep[e.g.,][and references therein]{kom+13}.  
Studies of Galactic pulsar wind nebulae (often referred to as plerions) suggest that a significant fraction 
of the spin-down energy is dissipated at the termination shock, where it converts into the non-thermal energy 
of accelerated particles, which are primarily electrons and positrons~\citep{rg74,kc84mhd}.  
Embryonic nebular emission has also been of interest in the literature of gamma-ray bursts (GRBs)~
\citep[e.g.,][]{Usov_1992,Thompson_1994,Blackman_Yi_1998,dl98,Zhang_Meszaros_2001} including the ``supranova'' 
model~\citep[e.g.,][]{1998ApJ...507L..45V,2003ApJ...583..379I}, energetic SNe such as super-luminous SNe (SLSNe)~
\citep[e.g.,][]{Thompson_et_al_2004,Maeda_et_al_2007,Kasen_Bildsten_2010,Woosley_2010,2015ApJ...805...82M,
Kashiyama+16}, and kilonovae/macronovae~\citep[e.g.,][]{2015ApJ...802..119K,2015ApJ...807..163G,2016ApJ...818..104K}.
An efficient conversion of the rotation energy into particle energy is required for the pulsar-driven SN 
model to explain these energetic SNe.  

In this work, we study the consequences of an impulsive burst that occurs inside a wind bubble and SN ejecta.  
Multi-wavelength observations are crucial for distinguishing among the various models, and here we show that 
the existence of wind bubbles leads to interesting implications for the observations.   
We investigate three representative cases as examples, (a) the magnetar scenario, (b) the rapidly-rotating 
neutron star (RNS) scenario, and (c) the magnetized white dwarf (MWD) scenario. 
In Section~2 we outline our goals and we set up a  model of the nebula emission based on the theoretical 
modeling of Galactic pulsar wind nebulae.   
Then, independently of the details of the emission mechanisms, we consider the fate of high-energy gamma-rays 
and radio waves propagating in the nebula and in the baryonic ejecta, and show that the escape of gamma-rays and 
radio waves is possible for NSs with ages of $T\gtrsim10-100$~yr. 
In Section~3, we discuss the impact of a highly relativistic outflow that originates from an impulsive 
magnetic dissipation event. 
The decelerating relativistic flow may boost the pre-existing non-thermal particles in the nebula, and 
high-energy gamma-ray flashes (HEGFs) are expected in the GeV-TeV range.  
We also argue that possible dissipation at the reverse shock leads to lower-energy gamma-ray emission in the MeV-GeV range. 
In Section~4, we discuss the possibility of a subsequent afterglow emission, and show that interactions 
with the dense baryonic ejecta may lead to detectable radio emission for nearby bursts, especially in the magnetar scenario.   
In Section~5, we discuss possible constraints on the magnetar giant flare scenario, as well as some issues 
arising in merger scenarios.  
Throughout this work, we use the notation $Q={10}^xQ_x$ in CGS unit unless noted otherwise.

\begin{figure}
\includegraphics[width=\linewidth]{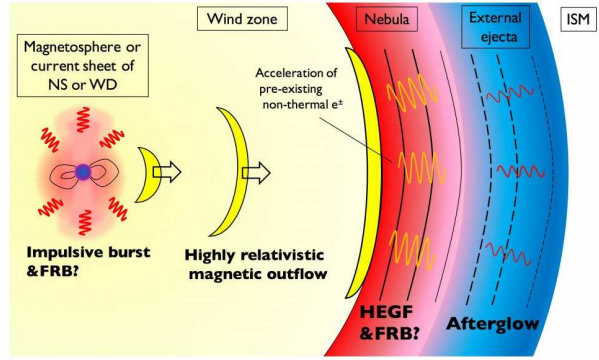}
\caption{Schematic picture of the burst-in-bubble model.  
\label{fig:schematic_pic}
}
\end{figure}

\section{Setup: properties of wind bubbles}
\subsection{Simplified nebula dynamics}
\begin{figure*}
\includegraphics[width=0.32\linewidth]{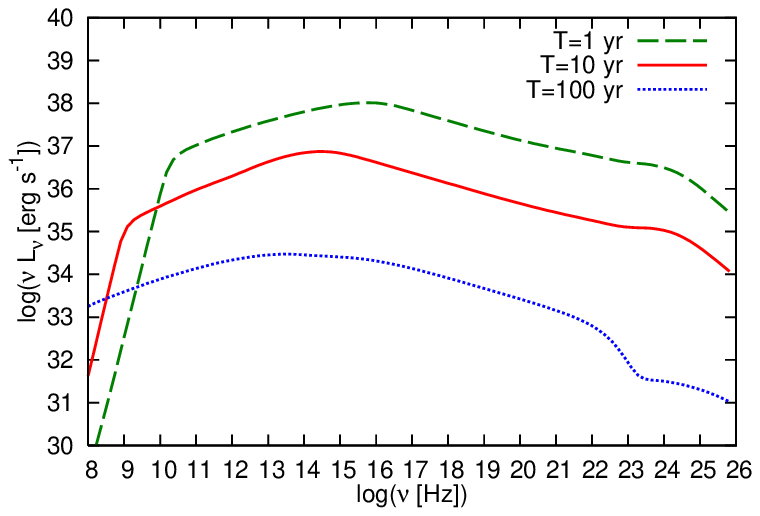}\hfill
\includegraphics[width=0.32\linewidth]{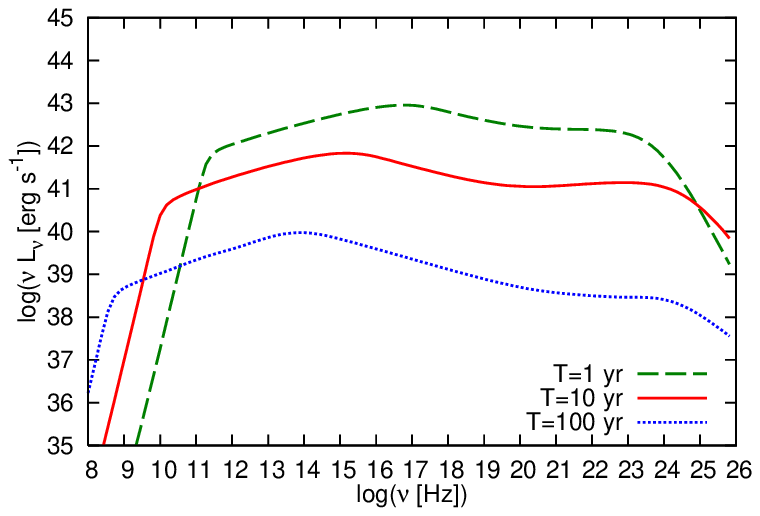}\hfill
\includegraphics[width=0.32\linewidth]{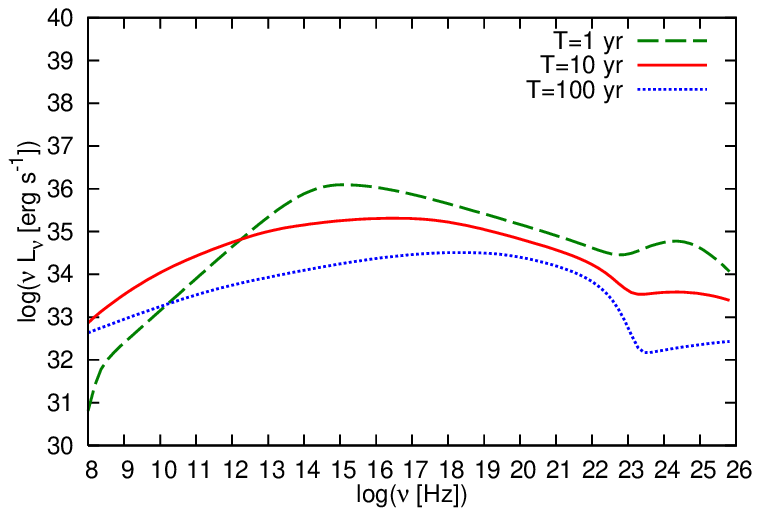}
\caption{Spectra of a very young nebula in the magnetar scenario (left panel), RNS scenario (middle panel), 
and MWD scenario (right panel).  Numerical calculations are performed based on \citet[][]{2015ApJ...805...82M}. 
Effects of attenuation in the ejecta are not shown (see text).  The microphysical parameters are set to $\epsilon_B=0.01$, 
$\epsilon_e=1-\epsilon_B$, $\gamma_b={10}^5$, $q_1=1.5$, $q_2=2.5$, and the parameters related to the dynamics are given in the text.  
\label{fig2}
}
\end{figure*}

We consider young nebulae that are formed by the spin-down activity of NSs (or WDs) embedded in the SN ejecta 
(or merger ejecta).  Such a scenario is naturally expected in many models involving nascent pulsars and 
MWDs~\footnote{Of course, there are exceptions. For example, if NSs or magnetars are formed by 
the accretion-induced collapse, we do not have to expect the existence of the dense SN ejecta.}, for not 
only FRBs but also SLSNe, GRBs, kilonovae/macronovae, high-energy neutrino emitters~\citep[e.g.,][]{mur+09,fan+14}, 
and even TeV cosmic-ray electron-positron factories~\citep{Kashiyama+11}, although reference parameters vary among the 
scenarios (see Fig.~\ref{fig:schematic_pic}). 
The nebular emission is powered by the rotation energy, whose energy budget is given by
\begin{eqnarray}
{\mathcal E}_{\rm rot}&=&\frac{1}{2}I\Omega_i^2\nonumber\\
&\simeq&
\left\{ \begin{array}{ll} 
1.9\times10^{47}~{\rm erg}~M_{*,\rm Ch}R_{*,6}^2P_{i,-0.5}^{-2}
& \mbox{(magnetar)}\\
1.9\times10^{52}~{\rm erg}~M_{*,\rm Ch}R_{*,6}^2P_{i,-3}^{-2}
& \mbox{(RNS)}\\
3.5\times10^{49}~{\rm erg}~M_{*,1M_\odot}R_{*,8.7}^2P_{i,1}^{-2}
& \mbox{(MWD)}\\
\end{array} \right.
\end{eqnarray}
where $I\approx0.35M_*R_*^2$ is the moment of inertia
Hereafter, we consider $M_*\approx M_{\rm Ch}=1.4~M_\odot$ (where $M_{\rm Ch}$ is the Chandrasekhar mass) 
and $R_*\approx10^6$~cm for NSs, 
whereas we adopt $M_*\approx1.0~M_\odot$ and $R_*\approx10^{8.7}$~cm for WDs.
(For the rest of our discussion in Section~2.1, we do not show the explicit dependence on $R_*$. )
Note that $\Omega_i=2\pi/P_i$ is the rotation frequency at the birth of the compact object.
For Galactic pulsars, the typical value of the initial period at the birth is estimated to be 
$P_i\sim300$~ms~\citep{fk06}, which is used for the magnetar scenario.  
Note that the period becomes $P(T)\gtrsim5$~s after 1000~yr.  
We consider $P_i={10}^{-3}$~s in the RNS scenario, and $P_i=10$~s in the MWD scenario. 

Based on recent magnetohydrodynamics simulations~\citep{Gruzinov_2005,Spitkovsky_2006,Tchekohovskoy_et_al_2013}, 
the initial spin-down power is 
\begin{eqnarray}
L_{{\rm sd},i}&\approx&\frac{B_*^2R_*^6{(2\pi/P_i)}^4}{4c^3}(1+\sin^2\chi)\nonumber\\
&\simeq&
\left\{ \begin{array}{ll} 
2.4\times10^{39}~{\rm erg}~{\rm s}^{-1}~B_{*,15}^2P_{i,-0.5}^{-4}
& \mbox{(magnetar)}\\
2.4\times10^{44}~{\rm erg}~{\rm s}^{-1}~B_{*,12.5}^2P_{i,-3}^{-4}
& \mbox{(RNS)}\\
3.8\times10^{37}~{\rm erg}~{\rm s}^{-1}~B_{*,9}^2P_{i,1}^{-4}
& \mbox{(MWD)}\\
\end{array} \right.
\end{eqnarray}
where $\chi$ is the angle between the rotation and magnetic axes, and numerical values are obtained with 
$<\sin^2\chi>=2/3$.  
Nominal magnetic field strengths are set to $B_*\sim10^{15}$~G in the magnetar scenario, 
$B_*\sim10^{12.5}$~G in the RNS scenario, and $B_*\sim10^{9}$~G in the MWD scenario.  
Note that the magnetar scenario may include the magnetar hyper-flare model for FRBs~\citep{Popov&Postnov10,Lyubarsky14}, 
although the association of soft gamma-ray emission depends on its dissipation mechanisms.  
The RNS scenario may include the blitzar model and other models motivated by giant pulses from 
the Crab pulsar~\citep{2014A&A...562A.137F,Cordes&Wasserman16}.  
The MWD scenario is a new possibility considered in this work.  Note that this MWD scenario is different 
from the WD-WD merger model of FRBs~\citep{Kashiyama+13} in the sense that the magnetic dissipation occurs 
much later than the MWD birth~\footnote{In \citet{Kashiyama+13}, FRB emission is expected to occur in a 
polar region, where the ambient density could be small enough.}.
In the MHD approximation, the spin-down power is significant even for aligned rotators 
with $\chi=0$, where the spin-down time becomes zero in the dipole formula 
(so that quantitative results in some models such as the blitzar model can be affected).
The corresponding spin-down time is given by
\begin{eqnarray}
T_{\rm sd}&=&\frac{3P_i^2Ic^3}{10\pi^2B_*^2R_*^6}\nonumber\\
&\simeq&
\left\{ \begin{array}{ll} 
2.5~{\rm yr}~B_{*,15}^{-2}P_{i,-0.5}^{2}
& \mbox{(magnetar)}\\
2.5~{\rm yr}~B_{*,12.5}^{-2}P_{i,-3}^{2}
& \mbox{(RNS)}\\
2.9\times10^{4}~{\rm yr}~B_{*,9}^{-2}P_{i,1}^{2}
& \mbox{(MWD)}\\
\end{array} \right.
\end{eqnarray}
After $T>T_{\rm sd}$, the rotation period increases as $P(T)\propto T^{1/2}$ and the spin-down power at 
the age $T$ decreases as $L_{\rm sd}(T)\propto T^{-2}$. 
In reality, the braking indices observed for Galactic pulsars differ from the theoretical value, 
but we use this formula for simplicity.

Next, we consider the baryonic ejecta outside the nebula. 
In the magnetar and RNS scenarios, we typically expect that the wind bubble is surrounded by the SN ejecta and 
we assume $M_{\rm ext}\sim3~M_{\odot}$ and $V_{\rm ext}\sim5000~{\rm km}~{\rm s}^{-1}$, motivated by 
applications to luminous SNe including SLSNe~\citep{Kashiyama+16}. Larger masses lead to more compact nebulae, 
where effects on the attenuation of non-thermal emission would become larger~\citep{2015ApJ...805...82M}. 
In the MWD scenario, we assume that MWDs are formed by merger events, where the ejecta has 
$\sim{10}^{-3}~M_{\odot}$ and $V_{\rm ext}\sim10000~{\rm km}~{\rm s}^{-1}$~\citep{2013ApJ...773..136J}.
In this work, we denote the SN or merger ejecta as the ``external'' baryonic ejecta, which is expected to freely 
expand until the deceleration radius $R_{\rm ST}$.  This radius is expressed as
\begin{eqnarray}
R_{\rm ST}&=&{\left(\frac{3M_{\rm ext}}{4\pi n_{\rm ism}m_p}\right)}^{1/3}\nonumber\\
&\simeq&
\left\{ \begin{array}{ll} 
9.6\times{10}^{18}~{\rm cm}~M_{\rm ext,0.5}^{1/3}n_{\rm ism}^{-1/3}
& \mbox{(SN)}\\
6.6\times{10}^{17}~{\rm cm}~M_{\rm ext,-3}^{1/3}n_{\rm ism}^{-1/3}
& \mbox{(merger)}\\
\end{array} \right.
\end{eqnarray}
The deceleration time is
\begin{equation}
T_{\rm ST}\simeq
\left\{ \begin{array}{ll} 
5.4\times{10}^{2}~{\rm yr}~V_{\rm ext,8.75}^{-1}M_{\rm ext,0.5}^{1/3}n_{\rm ism}^{-1/3}
& \mbox{(magnetar)}\\
1.2\times{10}^{2}~{\rm yr}~P_{i,-3}M_{\rm ext,0.5}^{5/6}n_{\rm ism}^{-1/3}
& \mbox{(RNS)}\\
21~{\rm yr}~V_{\rm ext,9}^{-1}M_{\rm ext,-3}^{1/3}n_{\rm ism}^{-1/3}
& \mbox{(MWD)}\\
\end{array} \right.
\end{equation}
After the deceleration time, the nebula may be disrupted by the reverse shock~\citep{2001ApJ...563..806B}.  
Thus, we consider very young nebulae with $T\lesssim100-1000$~yr, although magnetic dissipation itself 
could happen in middle-age nebulae. The ejecta radius $R_{\rm ext}\approx V_{\rm ext}T$ is estimated to be
\begin{eqnarray}
R_{\rm ext}\simeq
\left\{ \begin{array}{ll} 
1.8\times{10}^{17}~{\rm cm}~V_{\rm ext,8.75}T_{\rm 10yr}
& \mbox{(magnetar)}\\
7.8\times{10}^{17}~{\rm cm}~P_{i,-3}^{-1}M_{\rm ext,0.5}^{-1/2}T_{\rm 10yr}
& \mbox{(RNS)}\\
3.2\times{10}^{17}~{\rm cm}~V_{\rm ext,9}T_{\rm 10yr}
& \mbox{(MWD)}\\
\end{array} \right.
\end{eqnarray}
Note that the external ejecta velocity $V_{\rm ext}$ is governed by the rotation energy if 
${\mathcal E}_{\rm rot}$ exceeds the original explosion energy ${\mathcal E}_{\rm exp}\sim{10}^{51}$~erg, 
and the expression for $T>T_{\rm sd}$ is shown here.  
The corresponding nucleon density is estimated to be 
\begin{equation}
n_{\rm ext}\simeq
\left\{ \begin{array}{ll} 
1.6\times{10}^{5}~{\rm cm}^{-3}~M_{\rm ext,0.5}V_{\rm ext,8.75}^{-3}T_{\rm 10yr}^{-3}
& \mbox{(magnetar)}\\
1.9\times{10}^{3}~{\rm cm}^{-3}~P_{i,-3}^{3}M_{\rm ext,0.5}^{5/2}T_{\rm 10yr}^{-3}
& \mbox{(RNS)}\\
9.0~{\rm cm}^{-3}~M_{\rm ext,-3}V_{\rm ext,9}^{-3}T_{\rm 10yr}^{-3}
& \mbox{(MWD)}\\
\end{array} \right.
\end{equation}
Note that the free electron density would be significantly smaller since the ionization degree of the 
ejecta is expected to be low especially deep inside the ejecta.  
The typical SN ejecta becomes essentially 
neutral a few years after the SN explosion~\citep{1984ApJ...287..282H}.  However, especially at early times, 
the column density of free electrons may reach $\sim0.1-1000~{\rm pc}~{\rm cm}^{-3}$, which could contribute 
to the DM. 

As noted above, the characteristic ejecta velocity is not affected by a central remnant for ${\mathcal E}_{\rm rot}<{\mathcal E}_{\rm exp}$.  
In this case, the nebular radius is given by~\citep{che77}
\begin{equation}
R_{\rm nb}\approx{\left(\frac{125V_{\rm ext}^3L_{{\rm sd},i}}{99M_{\rm ext}}\right)}^{1/5}
\left\{ \begin{array}{ll} 
T^{6/5}
& (T<T_{\rm sd})\\
T_{\rm sd}^{6/5}{(T/T_{\rm sd})}
& (T>T_{\rm sd})\\
\end{array} \right.
\end{equation}
In the magnetar scenario, we have 
\begin{equation}
R_{\rm nb}\simeq
\left\{ \begin{array}{ll} 
2.5\times{10}^{15}~{\rm cm}~B_{*,15}^{2/5}P_{i,-0.5}^{-4/5}M_{\rm ext,0.5}^{-1/5}V_{\rm ext,8.75}^{3/5}T_{\rm yr}^{6/5}\\
2.9\times{10}^{16}~{\rm cm}~P_{i,-0.5}^{-2/5}M_{\rm ext,0.5}^{-1/5}V_{\rm ext,8.75}^{3/5}T_{\rm 10yr}
\end{array} \right.
\end{equation}
whereas in the MWD scenario we have
\begin{eqnarray}
R_{\rm nb}\simeq
1.2\times{10}^{17}~{\rm cm}~B_{*,9}^{2/5}P_{i,1}^{-4/5}M_{\rm ext,-3}^{-1/5}V_{\rm ext,9}^{3/5}T_{\rm 10yr}^{6/5}.
\end{eqnarray}

When the rotation energy exceeds the original explosion energy, the ejecta dynamics is modified by the 
central remnant. Then, we have~\citep{2005ApJ...619..839C} 
\begin{equation}
R_{\rm nb}\approx{\left(\frac{8L_{{\rm sd},i}}{15M_{\rm ext}}\right)}^{1/2}
\left\{ \begin{array}{ll} 
T^{3/2}
& (T<T_{\rm sd})\\
T_{\rm sd}^{3/2}{(T/T_{\rm sd})}
& (T>T_{\rm sd})\\
\end{array} \right.
\end{equation}
The nebular emission is assumed to come from a uniform sphere expanding at a constant velocity.  
This assumption is the easiest way to include the expansion of the nebula, although the realistic 
behavior of the expansion is more complicated.  Then, we have 
\begin{equation}
R_{\rm nb}\simeq
\left\{ \begin{array}{ll} 
2.5\times{10}^{16}~{\rm cm}~B_{*,12.5}P_{i,-3}^{-2}M_{\rm ext,0.5}^{-1/2}T_{\rm yr}^{3/2}\\
4.0\times{10}^{17}~{\rm cm}~P_{i,-3}^{-1}M_{\rm ext,0.5}^{-1/2}T_{\rm 10yr}
\end{array} \right.
\end{equation}
in the RNS scenario.  We see that the nebular radius is expected to be $R_{\rm nb}\sim{10}^{16}-{10}^{18}$~cm 
for $T\sim10-100$~yr.

The nebular mass is uncertain, but a reasonable lower limit can be placed using the Goldreich-Julian density~\citep{gj69}.    
Using values at the co-rotation radius, the mass outflow rate at $T<T_{\rm sd}$ is estimated to be 
\begin{equation}
\mdot_{\rm GJ}\simeq
\left\{ \begin{array}{ll} 
2.5\times10^{-18}~M_\odot~{\rm s}^{-1}~\mu_{\pm,6}B_{*,15}P_{i,0.5}^{-2}
& \mbox{(magnetar)}\\
7.9\times10^{-14}~M_\odot~{\rm s}^{-1}~\mu_{\pm,6}B_{*,12.5}P_{i,-3}^{-2}
& \mbox{(RNS)}\\
3.1\times10^{-17}~M_\odot~{\rm s}^{-1}~\mu_{\pm,6}B_{*,9}P_{i,1}^{-2}
& \mbox{(MWD)}
\end{array} \right.
\end{equation}
where $\mu_\pm$ is the pair multiplicity (defined as the ratio of the pair density to the Goldreich-Julian density)
and the wind is believed to be dominated by electron-positron pairs. 
Then, for $\mu_\pm\sim10^5-10^6$, the mass accumulated during $T$ would be
\begin{equation}
M_{\rm nb}\gtrsim{10}^{-9}-{10}^{-5}~M_{\odot}. 
\end{equation}
In this work, we use $M_{\rm nb}\sim{10}^{-7}~M_\odot$ as a reference value, which is comparable to 
the ejecta mass suggested to explain afterglow emission following the 2004 giant flare of SGR 1806-20
~\citep{2005Natur.434.1104G,2006ApJ...638..391G}. The nebular density is then estimated to be 
\begin{equation}
n_{\rm nb}\simeq
2.1\times{10}^{3}~{\rm cm}^{-3}~M_{\rm nb,-7}P_{i,-0.5}^{6/5}M_{\rm ext,0.5}^{3/5}V_{\rm ext,8.75}^{-9/5}T_{\rm 10yr}^{-3}
\end{equation}
in the magnetar scenario, 
\begin{equation}
n_{\rm nb}\simeq
8.0\times{10}^{-1}~{\rm cm}^{-3}~M_{\rm nb,-7}P_{i,-0.5}^{3}M_{\rm ext,0.5}^{3/2}T_{\rm 10yr}^{-3}
\end{equation}
in the RNS scenario, and
\begin{eqnarray}
n_{\rm nb}&\simeq&
1.7\times{10}^{-2}~{\rm cm}^{-3}~M_{\rm nb,-7}B_{*,9}^{-6/5}P_{i,1}^{12/5}\nonumber\\
&\times&M_{\rm ext,-3}^{3/5}V_{\rm ext,9}^{-9/5}T_{\rm 10yr}^{-18/5}
\end{eqnarray}
in the MWD scenario, respectively. 
Note that the column density may reach $n_{\rm nb}R_{\rm nb}\sim0.001-100~{\rm pc}~{\rm cm}^{-3}$, which 
could contribute to the DM. 

Note that the pair multiplicity is very uncertain and it is difficult to give a plausible theoretical value from 
first principles.  As a result, its effect on the DM may be subject to a significant uncertainty.
although results on the spectra are more insensitive~\citep[cf.][]{2013PTEP.2013l3E01T}.
On the other hand, our calculations for HEGFs and their afterglows, which are shown in Section~3 and 4, 
are largely unaffected by this.

\subsection{Quasi-steady nebular emission}
Next, we model the emission of the nebula.  
Following \cite{2015ApJ...805...82M}, we assume that the nebular 
emission mechanism and the parameters are similar to that for Galactic pulsar wind nebulae~\citep[e.g.,][and 
references therein]{gs06,tt10}.  For electrons and positrons, we consider a broken power-law injection spectrum,
\begin{equation}\label{eq:inj}
\frac{d \dot{n}_e^{\rm nb}}{d\gamma_e}
\propto
\left\{ \begin{array}{ll} 
\gamma_e^{-q_1}
& \mbox{($\gamma_m\leq\gamma_e\leq\gamma_b$)}\\
\gamma_e^{-q_2}
& \mbox{($\gamma_b<\gamma_e\leq\gamma_M$)}
\end{array} \right.
\end{equation} 
where $q_1\sim1-1.5(<2)$ and $q_2\sim2.5-3(>2)$ are low- and high-energy spectral indices, and 
$\gamma_b\sim{10}^{4}-{10}^6$ is the characteristic Lorentz factor of the accelerated leptons.  
A significant energy fraction ($\epsilon_e\sim1$) of $L_{\rm sd}$ is used for non-thermal lepton 
acceleration~\citep{rg74,kc84mhd,tt10,tt13}.  

At sufficiently late times, the dominant radiation mechanism is the synchrotron process. 
The magnetic field energy density in the nebula is parameterized as
\begin{equation}
U_B^{\rm nb}=\epsilon_B\frac{3\int dT\,\,L_{\rm sd}}{4\pi R_{\rm nb}^3},
\end{equation}
where $\epsilon_B\sim0.001-0.01$ is the magnetic energy fraction, and we use $\epsilon_B=0.01$ in this work.   
This value is based on results of detailed modeling of some of the known Galactic pulsar wind nebulae~\citep{tt10,tt13}.  
Although this assumption may not hold in early nebulae, it is reasonable and that our results in Sections~3 and 4 are not be much 
affected (see below). 
 
In the magnetar scenario, the nebular magnetic field strength is estimated to be
\begin{equation}
B_{\rm nb}\simeq2.1\times10^{-2}~{\rm G}~P_{i,-0.5}^{-2/5}M_{\rm ext,0.5}^{3/10}V_{\rm ext,8.75}^{-9/10}T_{\rm 10yr}^{-3/2}\epsilon_{B,-2}^{1/2}
\end{equation}
for $T>T_{\rm sd}$. Similarly, we have
\begin{equation}
B_{\rm nb}\simeq1.3\times10^{-1}~{\rm G}~P_{i,-3}^{1/2}M_{\rm ext,0.5}^{3/4}T_{\rm 10yr}^{-3/2}\epsilon_{B,-2}^{1/2}
\end{equation}
in the RNS scenario and
\begin{equation}
B_{\rm nb}\simeq6.5\times10^{-4}~{\rm G}~B_{*,9}^{2/5}P_{i,1}^{-4/5}M_{\rm ext,-3}^{3/10}V_{\rm ext,9}^{-9/10}T_{\rm 10yr}^{-13/10}\epsilon_{B,-2}^{1/2}
\end{equation}
in the MWD scenario, respectively. 
 
Taking the magnetar scenario as an example, the characteristic synchrotron frequency of the nebular 
emission is
\begin{eqnarray}
\nu_b&\approx&\frac{3}{4\pi}\gamma_b^2\frac{eB_{\rm nb}}{m_ec}\nonumber\\
&\simeq&9.0\times{10}^{14}~{\rm Hz}~\gamma_{b,5}^2P_{i,-0.5}^{-2/5}R_{*,6}^{2/5}\epsilon_{B,-2}^{1/2}\nonumber\\
&\times&M_{\rm ext,0.5}^{3/10}V_{\rm ext,8.5}^{-9/10}T_{\rm 10yr}^{-3/2}.
\end{eqnarray}
The cooling Lorentz factor of electrons is determined by $t_{\rm syn}=T$, which is
\begin{eqnarray}
\gamma_c&\approx&\frac{6\pi m_ec}{\sigma_T B_{\rm nb}^2T}\nonumber\\
&\simeq&5300~P_{i,-0.5}^{4/5}R_{*,6}^{-4/5}M_{\rm ext,0.5}^{-3/5}V_{\rm ext,8.5}^{9/5}T_{\rm 10yr}^{2}\epsilon_{B,-2}^{-1},
\end{eqnarray} 
and the corresponding cooling synchrotron frequency is 
\begin{eqnarray}
\nu_c&\approx&\frac{3}{4\pi}\gamma_c^2\frac{eB_{\rm nb}}{m_ec}\nonumber\\
&\simeq&2.6\times{10}^{12}~{\rm Hz}~P_{i,-0.5}^{6/5}R_{*,6}^{-6/5}\epsilon_{B,-2}^{-3/2}\nonumber\\
&\times&M_{\rm ext,0.5}^{-9/10}V_{\rm ext,8.5}^{27/10}T_{\rm 10yr}^{5/2}.
\end{eqnarray}
Note that the maximum synchrotron frequency is
\begin{eqnarray}
\nu_M\approx\frac{9}{4}\frac{m_ec^3}{2\pi e^2}\simeq3.8\times{10}^{22}~{\rm Hz}, 
\end{eqnarray}
which does not depend on $\epsilon_B$. This is obtained simply by equating half of the Larmor time
with the synchrotron cooling time~\citep[e.g.,][]{2014RPPh...77f6901B,2011MNRAS.414.2017K,2012MNRAS.424.2249K}. 

In the fast-cooling case ($\nu_c<\nu_b$), the synchrotron spectrum is expressed as
\begin{equation}
\nu L_{\nu}^{\rm nb}
\sim\frac{\epsilon_eL_{\rm sd}}{2{\mathcal R}_b}
\left\{ \begin{array}{ll} 
{(\nu/\nu_b)}^{(2-q_1)/2}
& \mbox{($\nu_c \leq \nu \leq \nu_b$)}\\
{(\nu/\nu_b)}^{(2-q_2)/2}
& \mbox{($\nu_b <\nu \leq \nu_M$)}
\end{array} \right.
\end{equation}
where ${\mathcal R}_b\sim{(2-q_1)}^{-1}+{(q_2-2)}^{-1}\sim5$.  
The slow-cooling synchrotron spectrum ($\nu_b<\nu_c$) is written as
\begin{equation}
\nu L_\nu^{\rm nb}
\sim\frac{\epsilon_eL_{\rm sd}}{2{\mathcal R}_b}
\left\{ \begin{array}{ll} 
{(\nu_b/\nu_c)}^{(3-q_2)/2}{(\nu/\nu_b)}^{(3-q_1)/2}
& \mbox{($\nu \leq \nu_b$)}\\
{(\nu/\nu_c)}^{(3-q_2)/2}
& \mbox{($\nu_b \leq \nu \leq \nu_c$)}\\
{(\nu/\nu_c)}^{(2-q_2)/2}
& \mbox{($\nu_c <\nu \leq \nu_M$)}
\end{array} \right.
\end{equation}

We numerically calculate the spectra of the nebular emission following the method used in 
\citet[][]{2015ApJ...805...82M}. 
We include the synchrotron and inverse-Compton emission processes, and solve the time-dependent 
kinetic equations taking into account electromagnetic cascades. 
The cosmic microwave background and extragalactic background light are included as external radiation fields. 
Note that the SN or merger emission is important at early times from weeks to months~\citep{2015ApJ...805...82M,kot+13}, 
but such a very early phase is beyond the scope of this work.  
The synchrotron self-absorption (SSA) process and the Razin effect are also taken into account in the 
calculations.  The numerical results are shown in Fig.~2. 
See the Appendix for more details on light curves and a discussion on the detectability of emission from quasi-steady nebulae. 

Note that the nebular emission is relevant even if it is itself undetectable.  As we discuss in the next 
subsection, it can prevent high-energy gamma-rays and radio waves from escaping the nebula.

\subsection{Gamma-ray and radio attenuation in the nebula and ejecta}
\begin{figure*}
\includegraphics[width=0.32\linewidth]{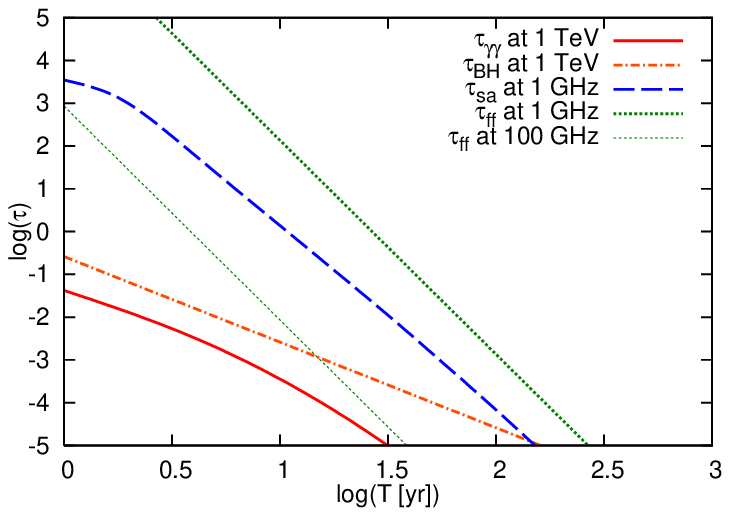}\hfill
\includegraphics[width=0.32\linewidth]{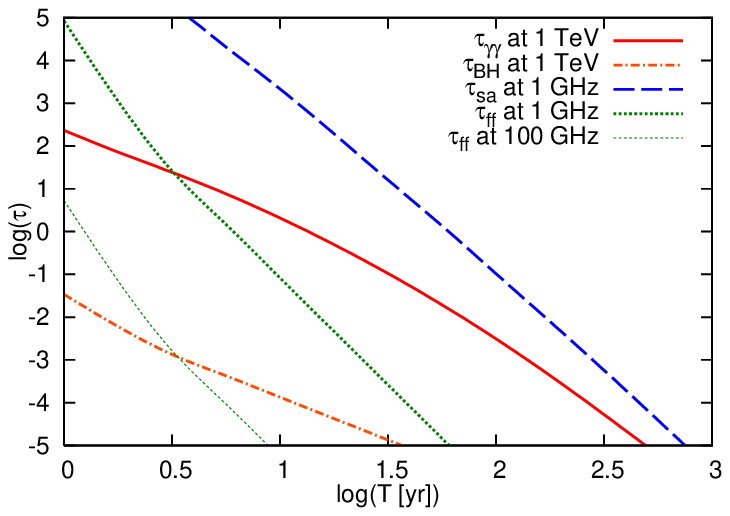}\hfill
\includegraphics[width=0.32\linewidth]{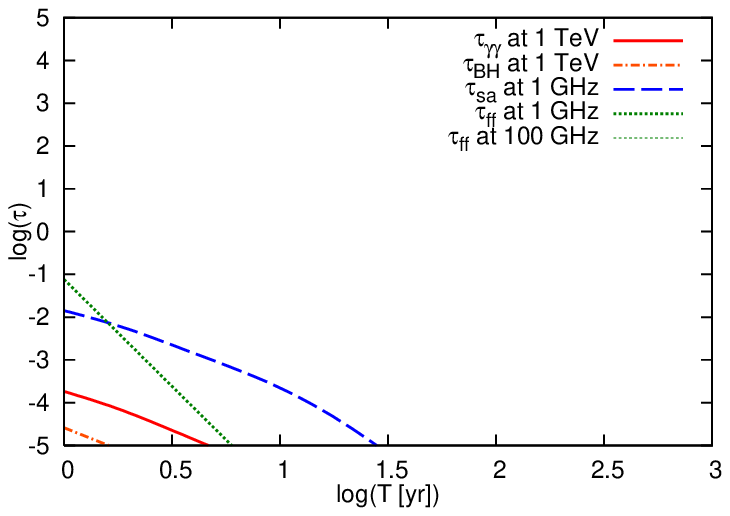}
\caption{Optical depths to the two-photon annihilation, Bethe-Heitler pair-production, SSA, and free-free 
absorption processes as a function of the age $T$, considering 1~TeV gamma-rays and radio waves at 1~GHz and 100~GHz as
examples.  Corresponding to Fig.~2, the values for the magnetar scenario (left panel), RNS scenario (middle panel), and MWD scenario 
(right panel) are shown. 
\label{fig3}
}
\end{figure*}

Based on the nebula model described above, we consider the fate of the radio and gamma-ray emission.   
As discussed in Section~3, HEGFs and/or FRBs may be generated inside the nebula, and it is natural 
to ask whether these emissions can escape or not from the nebula.  Our numerical results for various 
absorption processes in the three scenarios with nominal parameters are shown in Fig.~3 as a function of the age $T$. 
Here the calculations are performed based on the spectral energy distributions shown in Figs.~2. 

For gamma-rays, we consider the two-photon annihilation process ($\gamma\gamma\rightarrow e^-e^+$) in the nebula 
and the Bethe-Heitler process ($\gamma A\rightarrow e^-e^+A$) in the external ejecta. 
The optical depth to two-photon annihilation is calculated by
\begin{equation}
\tau_{\gamma \gamma}^{\rm nb}(\varepsilon_\gamma)=R_{\rm nb}\int d \nu \,\, n_{\nu}^{\rm nb} \int \frac{d \cos\theta}{2} \,\, (1-\cos\theta) \sigma_{\gamma \gamma}({\varepsilon_\gamma}, \cos\theta),
\end{equation}
where $n_\nu^{\rm nb}$ is the photon density in the nebula and $\theta$ is the angle between two photons. 
As shown in Fig.~3, we find that the gamma-ray escape of TeV photons is possible in all three 
scenarios~\footnote{Note that, as studied in \citet[][]{2015ApJ...805...82M}, the SN emission is relevant in embryonic 
nebulae ($T\lesssim1$~yr) and gamma-ray escape can be prevented by the two-photon annihilation process.}.
However, because of the large spin-down power, the escape is prevented until $T\sim10$~yr in the RNS scenario 
(see the middle panel of Fig.~3).  The Bethe-Heitler pair-production process is calculated following 
\cite{2015ApJ...805...82M}, and the resulting attenuation turns out to be more negligible even in the 
RNS and MWD scenarios.  

The escape of radio waves including FRB emission is more difficult. One of the relevant processes is the SSA 
process in the nebula~\citep[e.g.,][]{2016ApJ...819L..12Y}.  The SSA optical depth is calculated as
\begin{equation}\label{eq:taugammagamma}
\tau_{\rm sa}^{\rm nb}(\nu)=R_{\rm nb}\int d \gamma_e \,\, \frac{d n_e^{\rm nb}}{d\gamma_e}\sigma_{\rm sa}(\nu,\gamma_e),
\end{equation}
where the SSA cross section is~\citep[e.g.,][]{1991MNRAS.252..313G}
\begin{equation}
\sigma_{\rm sa}(\nu,\gamma_e)=\frac{1}{2m_e\nu^2\gamma_ep_e}\frac{\partial}{\partial \gamma_e}[\gamma_ep_ej_{\rm syn}(\nu,\gamma_e)],
\end{equation}
where $j_{\rm syn}$ is the synchrotron emissivity and $p_e$ is the electron momentum.  
The relevance of the SSA is already seen in Fig.~2, where an SSA cutoff appears at $\gtrsim1$~GHz at early times. 
The evolution of the SSA optical depth is shown in Fig.~3. In the magnetar scenario, the nebula is opaque against radio waves until $T\sim5$~yr, while in the RNS scenario, the GHz radio emission can get out only at times $T\gtrsim30$~yr.  However, as shown in Figs.~2 and 3, the escape of higher-frequency radio waves, e.g., at 100~GHz, is more readily possible. 

The free-free absorption in the external ejecta can be relevant, although it depends on the amount of free electrons, which is uncertain.  In Fig.~3, assuming a mean atomic number $\bar{A}=10$, $\bar{Z}=5$, and the singly ionized state~\footnote{In the earlier version of this paper, for the calculation of free-free absorption, the fully ionized state with $T_{\rm ext}=10^{2.5}$~K was actually assumed. Here more realistic parameters are used.} (implying the electron density $n_e^{\rm ext}\approx n_{\rm ext}\hat{\mu}_e^{-1}\sim n_{\rm ext}/\bar{A}$) with $T_{\rm ext}=10^4$~K~\citep{1984ApJ...287..282H}, we estimate the free-free absorption optical depth ($\tau_{\rm ff}$). Note that realistic values may be smaller since the external ejecta is expected to be essentially neutral at late times. As shown in Fig.~3, the free-free absorption can be the most relevant for the propagation of radio waves in the magnetar scenario, and the radio waves may be absorbed at $T\lesssim30-100$~yr. 
In the RNS scenario, due to the large spin-down power, the SSA process is expected to play a dominant role 
in the attenuation of radio waves.   
In the MWD scenario, the system is transparent to both radio and gamma-ray emission. 

We also calculate the DM due to the baryonic ejecta, and the results are shown in Fig.~4. 
In principle, it is possible for the baryonic ejecta to significantly contribute to the DM. 
However, interestingly, we find that the ejecta contribution may typically be limited to 
${\rm DM}\lesssim10-100~{\rm pc}~{\rm cm}^{-3}$, because large densities of ionized electrons 
also increase $\tau_{\rm ff}$.   

There are other plasma processes that could also be relevant. 
The plasma frequency is given by $\nu_{\rm pl}=(1/2\pi) \times {(4\pi n_{e}^{\rm ext}e^2/m_e)}^{1/2} \simeq 
9.0 \times 10^5 \ {\rm Hz} \ \sqrt{n_{e, 4}^{\rm ext}}$, below which electromagnetic waves cannot propagate 
in the medium.  One can see that the induced-Compton and induced-Raman scattering in the cold plasma can be 
neglected at $T\gtrsim30-100$~yr.  For example, the optical depth to the induced-Compton scattering in the 
external ejecta is estimated to be ~\citep[e.g.,][]{1978MNRAS.185..297W,2008ApJ...682.1443L}
\begin{eqnarray}
\tau_{\rm ind}^{\rm ext}(\nu)&\approx&\frac{3\sigma_{T}c n_e^{\rm ext} L_{\rm FRB}\Delta t}{32\pi^2m_e\nu^3R_{\rm ext}^2} \notag \\ 
&\simeq& 2.1 \times 10^{-6} \ L_{\rm FRB, 43} \Delta t_{-3} \nu_{9}^{-3} n_{e,4}^{\rm ext} R_{\rm ext, 17}^{-2}. \,\,\,\,\,
\end{eqnarray}
\begin{figure}
\includegraphics[width=0.9\linewidth]{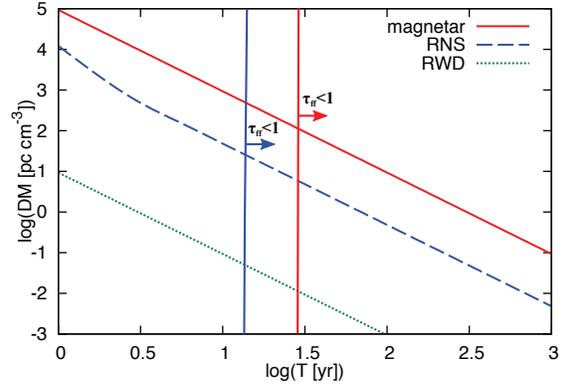}
\caption{Time evolution of the local DM due to the baryonic ejecta, for three scenarios considered in this work. 
\label{fig:schematic_pic}
}
\end{figure}

\section{High-Energy Gamma-Ray Flashes}
Pulsar winds are highly relativistic outflows that are initially Poynting-dominated~\citep{1969ApJ...158..727M}.  
An impulsive energy injection with an intrinsic duration $\delta t$, which may originate from magnetic dissipation 
in the magnetosphere~\citep[e.g.,][]{1995MNRAS.275..255T,2001ApJ...561..980T} 
or around current sheets in the wind zone~\citep[e.g.,][]{2010MNRAS.407.1926G,2016MNRAS.456.3282E}, or 
perhaps from the spin-down power, may lead to the formation of a highly relativistic outflow with Lorentz 
factors $\Gamma_0$.  Such a highly relativistic impulse is expected to initially propagate in the 
pre-existing pulsar wind, which is itself  also Poynting-dominated.  
We focus on cases of $\Gamma_0\gg1$, although the bulk acceleration depends on how many pairs are loaded in 
the flow as well as details of the magnetic dissipation~\footnote{There may be prompt emission associated 
with magnetic reconnections. In the magnetar scenario, it may be observed as giant flares (or short bursts 
in X-rays), as discussed in Section~5.}.   
The magnetic energy would be the most promising energy source in our model.  The magnetic energy tapped in 
the star is~\citep[cf.][]{1982ApJ...260..371K,Kashiyama+13,2014ApJ...780L..21Z}
\begin{eqnarray}
{\mathcal E}_{\rm mag}&\approx&\frac{1}{6}B_*^2R_*^3\nonumber\\
&=&
\left\{ \begin{array}{ll} 
1.7\times10^{48}~{\rm erg}~B_{c,15.5}^2R_{*,6}^3
& \mbox{(magnetar)}\\
1.7\times10^{43}~{\rm erg}~B_{c,13}^2R_{*,6}^3
& \mbox{(RNS)}\\
2.1\times10^{45}~{\rm erg}~B_{c,9.5}^2R_{*,8.7}^3
& \mbox{(MWD)}
\end{array} \right.
\end{eqnarray}
where $B_c$ is the internal magnetic field, which could be larger than the surface magnetic field $B_*$.  
Note also that, in principle, the spin-down power $L_{\rm sd}$ can be relevant as an alternative energy source in 
a certain model involving RNSs~\citep[e.g.,][for NS-NS mergers]{2013PASJ...65L..12T}.

Hereafter, we use $t$ as the elapsed time after a burst.  The distance between a compact remnant and its 
nebula is denoted by $r_0$, which can be smaller than $R_{\rm nb}$. 
Note that the central remnant may receive a kick by, e.g., the SN explosion.  In the NS case, the typical 
kick velocity is $V_k\sim500~{\rm km}~{\rm s}^{-1}$.  The distance the NS travels is $R_k\approx 
V_k T\simeq1.6\times{10}^{17}~{\rm cm}~(V_k/500~{\rm km}~{\rm s}^{-1})T_{10{\rm yr}}$, which can be comparable 
to $R_{\rm nb}$, and $r_0$ is given by $r_0=R_{\rm nb}-R_k$ for $R_k<R_{\rm nb}$.

\subsection{GeV-TeV flashes from a forward shock}
An impulsive flow with $\Gamma_0\sim{10}^4-{10}^6$ is decelerated as soon as it reaches the nebula.  
The deceleration is quick since the deceleration radius 
$r_\Gamma\approx{[3{\mathcal E}/(4\pi n_{\rm nb}m_ec^2\Gamma_0^2)]}^{1/3}$ is typically much smaller than $r_0$. 
For a sufficiently large energy injected into the nebula, the compressed nebula becomes relativistic.  
As suggested for the Crab pulsar flares~\citep[e.g.,][]{2014RPPh...77f6901B,2011MNRAS.414.2017K,2012MNRAS.424.2249K}, 
the Doppler-boosted nebular emission may generate photons whose energy is higher than the synchrotron energy limited 
by radiation losses, and a possible application to FRBs has also also discussed \citep{Lyubarsky14}.  
We assume that the shell width at $r_0$ is $c\Delta t$ in the observer frame, where $\Delta t\approx 
{\rm max}[\delta t, r_0/(2\Gamma_0^2c)]$ gives the duration of the HEGFs. 
The duration can be $\Delta t\approx\delta t\sim1$~ms (as in FRBs), but it may be as long as 
$\Delta t\approx r_0/(2\Gamma_0^2c)\simeq1700~{\rm s}~r_{0,16}\Gamma_{0,1}^{-2}$ (for GRB-flare-like Lorentz factors). 
The Lorentz factor of the shocked region $\Gamma$ can be determined by the pressure balance. 
For a magnetic piston or unmagnetized ejecta in the thick shell limit, $\Gamma$ is estimated
by ${\mathcal E}/(4\pi r^2\Gamma^2c\Delta t)\approx4\Gamma^2 p_{\rm nb}\xi$, 
where $p_{\rm nb}$ is the nebular pressure and $\xi<1$ is a pre-factor due to radiation energy losses.  
The Lorentz factor of the shocked region roughly becomes 
\begin{eqnarray}
\Gamma\sim
\left\{ \begin{array}{ll} 
1100~{\mathcal E}_{48}^{1/4}B_{*,15}^{1/2}{\Delta t}_{-3}^{-1/4}T_{\rm 10yr}^{1/2}
& \mbox{(magnetar)}\\
3.6~{\mathcal E}_{43}^{1/4}B_{*,12.5}^{1/2}{\Delta t}_{-3}^{-1/4}T_{\rm 10yr}^{1/2}
& \mbox{(RNS)}\\
280~{\mathcal E}_{45}^{1/4}B_{*,9}^{-1/2}P_{i,1}{\Delta t}_{-3}^{-1/4}
& \mbox{(MWD)}
\end{array} \right.
\end{eqnarray}
in the three scenarios.  Hereafter, we consider $\Gamma\sim10-1000$, and in this subsection, for demonstration
purposes, we use the reference parameters of the magnetar scenario. 

In young nebulae, the immediate upstream is filled with pre-accelerated particles with a typical Lorentz 
factor of $\gamma_b$ (for the original nebula).  If there is no cooling and the shell is thick enough for 
particles to be thermalized, the injection Lorentz factor (in the rest frame of the flow moving with $\Gamma$) 
is given by 
\begin{equation}
\gamma'_i\approx \Gamma \gamma_b=10^8\Gamma_3 \gamma_{b,5}, 
\end{equation}
and the corresponding synchrotron energy is given by
\begin{eqnarray}
\varepsilon_i&\approx&\frac{3}{4\pi}\Gamma\hbar{\gamma'}_i^2\frac{eB'}{m_ec}\nonumber\\
&\simeq&3.7~{\rm TeV}~\Gamma_3^4\gamma_{b,5}^2P_{i,-0.5}^{-2/5}R_{*,6}^{2/5}M_{\rm ext,0.5}^{3/10}V_{\rm ext,8.75}^{-9/10}T_{\rm 10yr}^{-3/2}.\,\,\,\,\,\,\,\,\,\,\,\,
\end{eqnarray}
As a conservative choice, we adopt the shock-compressed field value $B'\approx\Gamma B_{\rm nb}$ for the 
evaluation of synchrotron emission from the shocked nebula. 
The synchrotron cooling Lorentz factor (for the refreshed nebula) is given by 
\begin{equation}
\gamma'_c\approx8.4\times{10}^5~\Gamma_3^{-3}P_{i,-0.5}^{4/5}R_{*,6}^{-4/5}M_{\rm ext,0.5}^{-3/5}V_{\rm ext,8.75}^{9/5}T_{\rm 10yr}^{3}{\Delta t}_{-3}^{-1},
\end{equation}
and the corresponding synchrotron cooling energy is
\begin{equation}
\varepsilon_c\simeq0.26~{\rm GeV}~\Gamma_3^{-4}P_{i,-0.5}^{6/5}R_{*,6}^{-6/5}M_{\rm ext,0.5}^{-9/10}V_{\rm ext,8.75}^{27/10}T_{\rm 10yr}^{9/2}{\Delta t}_{-3}^{-2}.
\end{equation}

Note that the Lorentz factor of accelerated leptons is limited by $r'_L/(2c)=6\pi m_ec/(\gamma'\sigma_T{B'}^2)$ 
and $r'_L/(2c)=l'/c\approx\Gamma\Delta t$.  The former condition gives
\begin{eqnarray}
\gamma'_{M1}\approx{\left(\frac{3m_e^2c^4}{2B'e^3}\right)}^{1/2}&\simeq&2.1\times{10}^7~\Gamma_3^{-1/2}P_{i,-0.5}^{1/5}R_{*,6}^{-1/5}\epsilon_{B,-2}^{-1/4}\nonumber\\
&\times&M_{\rm ext,0.5}^{-3/20}V_{\rm ext,8.75}^{9/20}T_{\rm 10yr}^{3/4},
\end{eqnarray}
and the latter condition gives
\begin{eqnarray}
\gamma'_{M2}\approx\left(\frac{2eB' l'}{m_ec^2}\right)&\simeq&7.5\times{10}^{8}~\Gamma_3^2P_{i,-0.5}^{-2/5}R_{*,6}^{2/5}\epsilon_{B,-2}^{1/2}\nonumber\\
&\times&M_{\rm ext,0.5}^{3/10}V_{\rm ext,8.75}^{-9/10}T_{\rm 10yr}^{-3/2}{\Delta t}_{-3}.
\end{eqnarray}
The corresponding maximum synchrotron energies are 
\begin{eqnarray}
\varepsilon_{M1}\simeq0.15~{\rm TeV}~\Gamma_3 
\end{eqnarray}
and 
\begin{eqnarray}
\varepsilon_{M2}&\simeq&2.1\times{10}^{2}~{\rm TeV}~\Gamma_3^6P_{i,-0.5}^{-6/5}R_{*,6}^{6/5}\epsilon_{B,-2}^{3/2}\nonumber\\
&\times&M_{\rm ext,0.5}^{9/10}V_{\rm ext,8.75}^{-27/10}T_{\rm 10yr}^{-9/2}{\Delta t}_{-3}^2,
\end{eqnarray}
and the maximum synchrotron energy is $\varepsilon_{M}={\rm min}[\varepsilon_{M1},\varepsilon_{M2}]$.
Note that $\varepsilon_{M1}$ does not depend on $\epsilon_B$ explicitly, so predictions for HEGFs may not be very
sensitive to the magnetic fields in the nebula. 

\begin{figure}
\includegraphics[width=\linewidth]{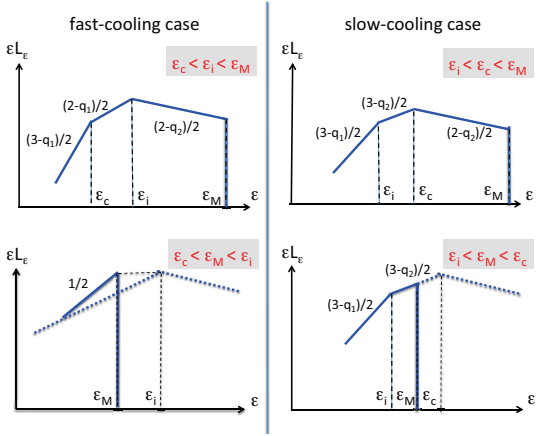}
\caption{Typical synchrotron spectra of HEGF emission.  The typical energy is expected to be at GeV-TeV energies for the forward shock emission. 
\label{fig:schematic_pic}
}
\end{figure}

In the conventional fast-cooling case ($\varepsilon_c<\varepsilon_i<\varepsilon_M$), the synchrotron spectrum 
is given by
\begin{equation}
\varepsilon L_{\varepsilon}^{\rm FS} \propto
\left\{ \begin{array}{ll} 
{(\varepsilon_c/\varepsilon_i)}^{(2-q_1)/2}{(\varepsilon/\varepsilon_c)}^{(3-q_1)/2}
& \mbox{($\varepsilon < \varepsilon_c$)}\\
{(\varepsilon/\varepsilon_i)}^{(2-q_1)/2}
& \mbox{($\varepsilon_c \leq \varepsilon \leq \varepsilon_i$)}\\
{(\varepsilon/\varepsilon_i)}^{(2-q_2)/2}
& \mbox{($\varepsilon_i < \varepsilon \leq \varepsilon_{M}$)}
\end{array} \right.
\end{equation}
However, for $\varepsilon_c<\varepsilon_M<\varepsilon_i$, most of the energy is released before the electrons 
and positrons are boosted to $\gamma'_i$.  In this situation, one expects that the lepton distribution would 
pile up around $\gamma'_M$, leading to a spectral bump around $\varepsilon_M$. 
\begin{equation}
\varepsilon L_{\varepsilon}^{\rm FS}\propto{(\varepsilon/\varepsilon_M)}^{1/2} \,\,\,\,\,\,\,\,\,\,\,\, \mbox{($\varepsilon \lesssim \varepsilon_M$)}
\end{equation}
The schematic picture is shown in Fig.~5. 
In the fast-cooling case, the radiation energy per logarithmic energy is 
\begin{equation}
(\varepsilon_M L_{\varepsilon_M}^{\rm FS}){\Delta t}\sim(\mathcal E/2{\mathcal R}_b)\sim{10}^{47}~{\rm erg}~{\mathcal E}_{48}.
\end{equation}
In the magnetar scenario, one can expect to be in  the fast-cooling case for $T\lesssim300$~yr (see below).  
Then, the TeV gamma-ray fluence from a HEGF is $\varepsilon^2 \phi_\gamma \sim8\times{10}^{-7}~
{\rm erg}~{\rm cm}^{-2}~{\mathcal E}_{48}d_{26}^{-2}$.  
On the other hand, the Fermi-LAT $5\sigma$ sensitivity at $\sim100$~GeV is $\sim{10}^{-4}~{\rm erg}~{\rm cm}^{-2}$, 
which is insufficient to detect such HEGFs. For this one would need larger gamma-ray detectors, such as the 
High-Altitude Water Cherenkov Observatory (HAWC), which has an effective area of $\sim100~{\rm m}^2$ at 100~GeV 
and $2\times{10}^4~{\rm m}^2$ at 1~TeV, respectively~\citep[][]{2013APh....50...26A}. 
The fluence sensitivity in the $0.1-1$~TeV range is $\sim{\rm a~few}\times{10}^{-7}~{\rm erg}~{\rm cm}^{-2}$, so 
HAWC may be able to detect a nearby HEGF within $\sim10-100$~Mpc if ${\mathcal E}\sim{10}^{46}-{10}^{48}$~erg. 
If the HEGF rate is $R_{\rm HEGF}\sim{10}^{4}~{\rm Gpc}^{-3}~{\rm yr}^{-1}$, the event rate is $\sim0.02-20~R_{\rm HEGF,4}~{\rm yr}^{-1}$. 
The future Cherenkov Telescope Array (CTA) is expected to have an effective area of $\sim10^6~{\rm m}^2$ at 1~TeV.  
The corresponding fluence sensitivity would be $\sim{10}^{-9}~{\rm erg}~{\rm cm}^{-2}$, implying that the
detection horizon of HEGFs is $\sim0.1-1$~Gpc. 
The CTA's Medium Size Telescopes of 12~m and 9.66~m will have a field of view of $8^\circ$, so the event rate 
in CTA's field of view may be $\sim0.006-6~R_{\rm HEGF,4}~{\rm yr}^{-1}$ with a duty cycle of 
10 percent~\footnote{For more detailed estimates of the event rate, see \citet{2012MNRAS.425..514K} that considers 
the detectability of GRBs at very high-energies.}.  
On the other hand, in the slow cooling regime, only a fraction of the particle energy is released as radiation, which is 
typically the case in the RNS and MWD scenarios.  For $\varepsilon_i<\varepsilon_c<\varepsilon_M$, we 
have the conventional slow-cooling spectrum, which is 
\begin{equation}
\varepsilon L_{\varepsilon}^{\rm FS} \propto
\left\{ \begin{array}{ll} 
{(\varepsilon_i/\varepsilon_c)}^{(3-q_2)/2}{(\varepsilon/\varepsilon_i)}^{(3-q_1)/2}
& \mbox{($\varepsilon < \varepsilon_i$)}\\
{(\varepsilon/\varepsilon_c)}^{(3-q_2)/2}
& \mbox{($\varepsilon_i \leq \varepsilon \leq \varepsilon_c$)}\\
{(\varepsilon/\varepsilon_c)}^{(2-q_2)/2}
& \mbox{($\varepsilon_c < \varepsilon \leq \varepsilon_M$)}
\end{array} \right.
\end{equation}
where the radiation energy is limited by
\begin{equation}
(\varepsilon_M L_{\varepsilon_M}^{\rm FS}){\Delta t}\sim(\mathcal E/2{\mathcal R}_b){(\varepsilon_c/\varepsilon_i)}^{(2-q_2)/2}.
\end{equation}
For $\varepsilon_i<\varepsilon_M<\varepsilon_c$, we obtain
\begin{equation}
\varepsilon L_{\varepsilon}^{\rm FS} \propto
\left\{ \begin{array}{ll} 
{(\varepsilon_i/\varepsilon_c)}^{(3-q_2)/2}{(\varepsilon/\varepsilon_i)}^{(3-q_1)/2}
& \mbox{($\varepsilon < \varepsilon_i$)}\\
{(\varepsilon/\varepsilon_c)}^{(3-q_2)/2}
& \mbox{($\varepsilon_i \leq \varepsilon \leq \varepsilon_M$)}\\
\end{array} \right.
\end{equation}
and 
\begin{equation}
(\varepsilon_M L_{\varepsilon_M}^{\rm FS}){\Delta t}\sim(\mathcal E/2{\mathcal R}_b){(\varepsilon_c/\varepsilon_i)}^{(2-q_2)/2}{(\varepsilon_{M2}/\varepsilon_c)}^{(3-q_2)/2}.
\end{equation}
In any case, the radiation energy is smaller than in the fast-cooling case, so it would be more difficult to 
be detected through HEGFs. 
Note that we do not consider cases of $\varepsilon_M<\varepsilon_i,\varepsilon_c$, where upstream particles 
with $\gamma'_i$ would be advected before they are thermalized or cooled by radiation. The radiation would be
very inefficient and the properties of such shocks are less clear.

A natural question concerns the conditions for the fast-cooling regime to be realized.  The critical 
Lorentz factor such that $\varepsilon_i=\varepsilon_c$ is given by
\begin{eqnarray}
\Gamma_{\rm cri}&\approx&300~\gamma_{b,5}^{-1/4}P_{i,-0.5}^{1/5}\epsilon_{B,-2}^{-1/4}\nonumber\\
&\times&M_{\rm ext,0.5}^{-3/20}V_{\rm ext,8.75}^{9/20}T_{\rm 10yr}^{3/4}{\Delta t}_{-3}^{-1/4}
\end{eqnarray}
in the magnetar scenario. Similarly, we have
\begin{eqnarray}
\Gamma_{\rm cri}&\approx&120~\gamma_{b,5}^{-1/4}P_{i,-3}^{-1/4}\epsilon_{B,-2}^{-1/4}\nonumber\\
&\times&M_{\rm ext,0.5}^{-3/8}T_{\rm 10yr}^{3/4}{\Delta t}_{-3}^{-1/4}
\end{eqnarray}
in the RNS scenario and
\begin{eqnarray}
\Gamma_{\rm cri}&\approx&1700~\gamma_{b,5}^{-1/4}B_{*,9}^{-1/5}P_{i,1}^{2/5}\epsilon_{B,-2}^{-1/4}\nonumber\\
&\times&M_{\rm ext,-3}^{-3/20}V_{\rm ext,9}^{9/20}T_{\rm 10yr}^{13/20}{\Delta t}_{-3}^{-1/4}
\end{eqnarray}
in the MWD scenario. 
The critical Lorentz factor at which $\varepsilon_M=\varepsilon_c$ is given by
\begin{eqnarray}
\Gamma_{\rm crM}&\approx&280~P_{i,-0.5}^{6/25}\epsilon_{B,-2}^{-3/10}\nonumber\\
&\times&M_{\rm ext,0.5}^{-9/50}V_{\rm ext,8.75}^{27/50}T_{\rm 10yr}^{9/10}{\Delta t}_{-3}^{-2/5}
\end{eqnarray}
in the magnetar scenario. Similarly, we have
\begin{eqnarray}
\Gamma_{\rm crM}&\approx&93~P_{i,-3}^{-3/10}\epsilon_{B,-2}^{-3/10}\nonumber\\
&\times&M_{\rm ext,0.5}^{-9/20}T_{\rm 10yr}^{9/10}{\Delta t}_{-3}^{-2/5}
\end{eqnarray}
in the RNS scenario and
\begin{eqnarray}
\Gamma_{\rm crM}&\approx&2300~B_{*,9}^{-6/25}P_{i,1}^{12/25}\epsilon_{B,-2}^{-3/10}\nonumber\\
&\times&M_{\rm ext,-3}^{-9/50}V_{\rm ext,9}^{27/50}T_{\rm 10yr}^{39/50}{\Delta t}_{-3}^{-2/5}
\end{eqnarray}
in the MWD scenario. These imply that the slow-cooling spectrum is typically expected in the RNS and MWD 
scenarios, and the radiation efficiency of HEGFs is not much larger. In the magnetar scenario, the efficient 
TeV emission is possible for $T\lesssim300$~yr. Hence, if a magnetic burst can occur during the lifetime 
(which is $\sim10^3-10^4$~yr in the magnetar scenario), we expect that strong HEGFs accompany a fraction 
of the bursts that occur in wind bubbles.

\subsection{MeV-GeV flashes from a reverse shock}
It is thought that the classical shock acceleration mechanism is inefficient for highly magnetized shocks~
\citep[e.g.,][]{2009ApJ...698.1523S}, so the non-thermal emission from the shocked wind would depend on 
the magnetization parameter of the impulsive flow ($\sigma_0$ at $r_0$).  
Here we discuss possible particle acceleration at the reverse shock which may occur especially if 
$\sigma_0\lesssim1$ or in the presence of magnetic reconnection.  We use the magnetar scenario as an example.

Assuming a flow with $\Gamma_0\gg1$, the relative Lorentz factor is estimated to be
\begin{equation}
\Gamma_{\rm rel}\approx\frac{\Gamma_0}{2\Gamma}=500~\Gamma_{0,6}\Gamma_3^{-1}. 
\end{equation}
As in the ordinary nebula case, we assume $\gamma'_b\approx\Gamma_{\rm rel}$. Then, the characteristic 
synchrotron energy in the observer frame becomes
\begin{eqnarray}
\varepsilon_b&\simeq&40~{\rm eV}~\Gamma_3^{-2}\Gamma_{0,6}^2{\mathcal E}_{48}^{1/2}P_{i,-0.5}^{2/5}R_{*,6}^{-6/5}\nonumber\\
&\times&M_{\rm ext,0.5}^{1/5}V_{\rm ext,8.75}^{-3/5}\epsilon_{B}^{1/2}T_{\rm 10yr}^{-1}{\Delta t}_{-3}^{-1/2}.
\end{eqnarray} 

On the other hand, the synchrotron cooling Lorentz factor (for the shocked wind) is given by 
\begin{equation}
\gamma'_c\approx5.0\times{10}^6~\Gamma_3{\mathcal E}_{48}^{-1}P_{i,-0.5}^{-4/5}R_{*,6}^{4/5}\epsilon_{B}^{-1}M_{\rm ext,0.5}^{-2/5}V_{\rm ext,8.75}^{6/5}T_{\rm 10yr}^{2},
\end{equation}
and the corresponding synchrotron cooling energy is
\begin{eqnarray}
\varepsilon_c&\simeq&3.8~{\rm GeV}~\Gamma_3^{2}{\mathcal E}_{48}^{-3/2}P_{i,-0.5}^{-6/5}R_{*,6}^{6/5}\epsilon_{B}^{-3/2}\nonumber\\
&\times&M_{\rm ext,0.5}^{-3/5}V_{\rm ext,8.75}^{9/5}T_{\rm 10yr}^{3}{\Delta t}_{-3}^{-1/2}.
\end{eqnarray}
The maximum synchrotron energy is
\begin{equation}
\varepsilon_M\simeq0.15~{\rm TeV}~\Gamma_3.
\end{equation}

The synchrotron spectrum is expected to be in the slow-cooling regime.  As in Eq.~(47), the released energy is limited by
\begin{eqnarray}
{\mathcal E}_{\rm RS}&\sim&3\times{10}^{46}~{\rm erg}~{\mathcal E}_{48}^{q_2-1}\Gamma_3^{4-2q_2}\Gamma_{0,6}^{q_2-2}\epsilon_{B}^{q_2-2}f_eP_{i,-0.5}^{4q_2/5-8/5}\nonumber\\
&\times&R_{*,6}^{8/5-4q_2/5}M_{\rm ext,0.5}^{2q_2/5-4/5}V_{\rm ext,8.75}^{12/5-6q_2/5}T_{\rm 10yr}^{4-2q_2}.
\end{eqnarray}
Thus, we expect that the forward shock emission is typically more relevant for the purpose of detecting 
flaring nebular emission. Note that the characteristic frequencies are sensitive to $\Gamma_{\rm rel}$, 
and smaller values lead to lower-energy gamma-ray emission that could be in the MeV range.

\section{Broadband afterglow emission}
In the previous section, we considered a burst with an explosion energy ${\mathcal E}\sim{10}^{43}-{10}^{48}$~erg. 
An energetic relativistic outflow triggered by the burst would sweep the nebula after it crosses the termination shock.  
Then, as the nebula is hot, the shock will disappear as the shocked flow becomes sub-sonic.  
The shell merges into the nebula, and would start expanding $\sqrt{3}r_{0}/c\simeq{10}^{6}~
{\rm s}~r_{0,16}$ after the burst.  Here we consider the refreshed forward shock formed as it interacts with 
the external baryonic ejecta.  We assume that the plasma in the shock vicinity is fully ionized by the external shock.  

To discuss the detectability, we use fiducial values of the magnetar scenario. 
However, the setup is generic and can be applied to giant flares of soft gamma-ray repeaters and merger scenarios. 
In the giant flare case, our model is actually analogous to its afterglow model~\citep{2005Natur.434.1104G}. 
In the merger case, a possible application could be considered if a long-lived magnetar exists. 

\begin{figure}
\includegraphics[width=\linewidth]{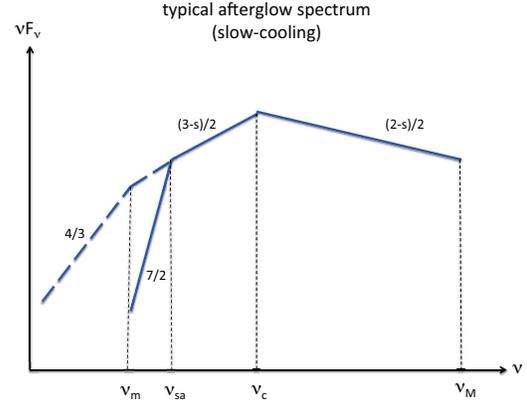}
\caption{A typical synchrotron spectrum of afterglow emission that follows a HEGF and FRB.   
\label{fig:schematic_pic}
}
\end{figure}

In the inelastic collision limit, assuming momentum and energy conservation, the merged shell (which we call 
the inner ejecta) has a Lorentz factor of 
\begin{equation}
\Gamma_{\rm ej}\approx \frac{1+{\mathcal E}/(M_{\rm nb}c^2)}{{[1+2{\mathcal E}/(M_{\rm nb}c^2)]}^{1/2}},
\end{equation} 
or $v_{\rm ej}\approx{\mathcal E}/(M_{\rm nb}c)$ in the non-relativistic limit. 
The inner ejecta with ${\mathcal E}\ll M_{\rm ext}c^2$ would be quickly decelerated (i.e. $r_{\rm dec}\ll r_0$) 
and the inner ejecta typically becomes non-relativistic at late times. Here, we focus on the refreshed forward 
shock emission after the inner ejecta starts the deceleration.  
Although the realistic behavior of the inner ejecta may be complicated at very early times, in this 
approximation, we may use the Sedov-Taylor solution.   Then, the shock velocity is given by
\begin{equation}
v_s\approx 5.2\times{10}^{9}~{\rm cm}~{\rm s}^{-1}~{\mathcal E}_{\rm AG,47.5}^{1/5}~n_{\rm ext,3}^{-1/5}t_{6}^{-3/5},
\end{equation}
where ${\mathcal E}_{\rm AG}$ is the kinetic energy of the inner ejecta and $n_{\rm ext}$ is the external ejecta density.  
The external shock radius is given by 
\begin{equation}
r_s\approx1.3\times{10}^{16}~{\rm cm}~{\mathcal E}_{\rm AG,47.5}^{1/5}~n_{\rm ext,3}^{-1/5}t_{6}^{2/5},
\end{equation}
which should be larger than $r_0$ for our approximation to be valid.  The downstream magnetic field is estimated 
to be 
\begin{eqnarray}
B_{\rm AG}&\approx&{(9\pi\epsilon_Bn_{\rm ext}m_pv_s^2)}^{1/2}\nonumber\\
&\simeq&0.11~{\rm G}~{\mathcal E}_{\rm AG,47.5}^{1/5}n_{\rm ext,3}^{3/10}\epsilon_{B,-2}^{1/2}t_{6}^{-3/5},\,\,\,\,\,\,
\end{eqnarray}
where $\epsilon_B\sim0.01$ is assumed in analogy to GRBs.  We assume that electrons are accelerated with 
a simple power law with $d{\mathcal N}_e/d\gamma_e\propto \gamma_e^{-s}$, 
where the minimum Lorentz factor of the accelerated electrons is
\begin{eqnarray}
\gamma_{m}&\approx&(\zeta_e/2)(m_p/m_e)(v_s^2/c^2)\nonumber\\
&\simeq&8.8~{\mathcal E}_{\rm AG,47.5}^{2/5}n_{\rm ext,3}^{-2/5}\zeta_{e,-0.5}t_{6}^{-6/5},\,\,\,
\end{eqnarray}
and the corresponding minimum synchrotron frequency is
\begin{eqnarray}
\nu_{m}&\approx&\frac{3}{4\pi}\gamma_{m}^2\frac{eB_{\rm AG}}{m_ec}\nonumber\\
&\simeq&3.7\times{10}^7~{\rm Hz}~{\mathcal E}_{\rm AG,47.5}n_{\rm ext,3}^{-1/2}\epsilon_{B,-2}^{1/2}\zeta_{e,-0.5}^2t_{6}^{-3},
\end{eqnarray}
where $\zeta_e$ represents the combination of the acceleration efficiency ($\epsilon_e$) and injection 
efficiency ($f_e$) at the forward shock.  The synchrotron cooling Lorentz factor is 
\begin{eqnarray}
\gamma_{c}&\approx&\frac{6\pi m_ec}{\sigma_TB_{\rm AG}^2 t}\nonumber\\
&\simeq&6.0\times{10}^4~{\mathcal E}_{\rm AG,47.5}^{-2/5}n_{\rm ext,3}^{-3/5}\epsilon_{B,-2}^{-1}t_{6}^{1/5},
\end{eqnarray}
and the synchrotron cooling frequency is
\begin{eqnarray}
\nu_{c}\simeq1.7\times{10}^{15}~{\rm Hz}~{\mathcal E}_{\rm AG,47.5}^{-3/5}n_{\rm ext,3}^{-9/10}\epsilon_{B,-2}^{-3/2}t_{6}^{-1/5}.
\end{eqnarray}
Thus, the synchrotron spectrum is typically expected to be in the slow-cooling regime.  
The maximum Lorentz factor is 
\begin{eqnarray}
\gamma_{M}&\approx&\sqrt{\frac{9\pi v_s^2e}{10\sigma_TB_{\rm AG}c^2}}\nonumber\\
&\simeq&2.3\times{10}^7~{\mathcal E}_{\rm AG,47.5}^{1/10}n_{\rm ext,3}^{-7/20}\epsilon_{B,-2}^{-1/4}t_{6}^{-3/10}, 
\end{eqnarray}
and the maximum synchrotron frequency is
\begin{eqnarray}
\nu_{M}\simeq2.6\times{10}^{20}~{\rm Hz}~{\mathcal E}_{\rm AG,47.5}^{2/5}n_{\rm ext,3}^{-2/5}t_{6}^{-6/5}. 
\end{eqnarray}
The synchrotron peak flux at $\nu_m$ or $\nu_c$ is
\begin{eqnarray}
F_\nu^{\rm max}&\approx&\frac{0.6f_e n_{\rm ext}r_s^3}{4\pi d^2}\frac{4\sqrt{3}\pi e^3B_{\rm AG}}{3m_ec^2}\nonumber\\
&\simeq&120~{\rm mJy}~{\mathcal E}_{\rm AG,47.5}^{4/5}n_{\rm ext}^{7/10}\epsilon_{B,-2}^{1/2}f_et_{6}^{3/5}d_{26}^{-2}.
\end{eqnarray}
In the slow-cooling regime, which is typically realized in all three scenarios, the synchrotron emission 
spectrum is
\begin{equation}
F_\nu^{\rm AG}\propto
\left\{ \begin{array}{ll} 
{(\nu_m/\nu_c)}^{(1-s)/2}{(\nu/\nu_m)}^{1/3}
& \mbox{($\nu \leq \nu_m$)}\\
{(\nu/\nu_c)}^{(1-s)/2}
& \mbox{($\nu_m <\nu \leq \nu_c$)}\\
{(\nu/\nu_c)}^{-s/2}
& \mbox{($\nu_c <\nu \leq \nu_M$)}
\end{array} \right.
\end{equation}
Noting that $\nu_m<\nu<\nu_c$, the radio flux is estimated to be
\begin{eqnarray}
F_\nu^{\rm AG}&\sim&12~{\rm mJy}~\nu_9^{1/2-s/2}{\mathcal E}_{\rm AG,47.5}^{3/10+s/2}n_{\rm ext,3}^{19/20-s/4}\nonumber\\
&\times&\epsilon_{B,-2}^{1/4+s/4}f_e{\zeta_{e,-0.5}}^{-1+s}t_{6}^{21/10-3s/2}d_{26}^{-2}, 
\end{eqnarray}
which implies that the radio emission may be detected by dedicated follow-up observations with present 
radio facilities such as the Very Large Array (VLA).  Note that the VLA flux sensitivity is $\sim0.03-0.1$~mJy 
at $t\gtrsim{10}^5$~s~\citep[e.g.,][]{2016arXiv160208492A}, so the radio signal may be detected up to 
$\sim10-100$~Mpc for ${\mathcal E}_{\rm AG}\sim{10}^{46.5}-{10}^{47.5}$~erg.  However, one should keep in 
mind that star-forming galaxies are bright in the radio band.  For example, the GHz radio flux of galaxies 
like the Milky Way is $\sim10$~mJy, so the detection of weak radio signals would be possible only for 
nearby events. 
Similarly, the continuum optical flux is estimated to be 
\begin{eqnarray}
F_\nu^{\rm AG}&\sim&0.75~{\mu{\rm Jy}}~\nu_{15}^{1/2-s/2}{\mathcal E}_{\rm AG,47.5}^{3/10+s/2}n_{\rm ext,3}^{19/20-s/4}\nonumber\\
&\times&\epsilon_{B,-2}^{1/4+s/4}f_e{\zeta_{e,-0.5}}^{-1+s}t_{6}^{21/10-3s/2}d_{26}^{-2},
\end{eqnarray}
which is in principle accessible with large optical telescopes such as Subaru and Keck. However, it is likely 
that such weak optical emission is masked by the host galaxy emission.
The X-ray energy flux at $\nu>\nu_c$ is written as 
\begin{eqnarray}
\nu F_\nu^{\rm AG}&\sim&2.5\times{10}^{-15}~{\rm erg}~{\rm cm}^{-2}~{\rm s}^{-1}~\nu_{18}^{1-s/2}{\mathcal E}_{\rm AG,47.5}^{s/2}\nonumber\\
&\times&n_{\rm ext,3}^{1/2-s/4}\epsilon_{B,-2}^{-1/2+s/4}f_e{(\zeta_e/0.4)}^{-1+s}t_{6}^{2-3s/2}d_{26}^{-2},\,\,\,\,\,\,\,\,\,\,
\end{eqnarray}
which could also be detectable by Chandra   
for very nearby bursts within $\sim3-30$~Mpc, the Chandra point-source flux sensitivity being 
$\sim4\times{10}^{-15}~{\rm erg}~{\rm cm}^{-2}~{\rm s}^{-1}$ with the integration time of $10^4$~s. 
Of course, the predicted fluxes depend on the values of ${\mathcal E}_{\rm AG}$. 

The SSA optical depth at $\nu>\nu_m$ in the slow-cooling regime is estimated by~
\citep[e.g.,][]{2004MNRAS.353..511P,mur+14isn},
\begin{equation}
\tau_{\rm sa}(\nu)\approx\frac{5 e n_{\rm ext}r_s}{3B_{\rm AG}\gamma_m^5}{\left(\frac{\nu}{\nu_m}\right)}^{-s/2-2}.
\end{equation}
The critical SSA frequency is determined by $\tau_{\rm sa}({\nu_{\rm sa}})=1$, which leads to
\begin{eqnarray}
\nu_{\rm sa}&\approx&2.5\times{10}^{9}~{\rm Hz}~{\mathcal E}_{\rm AG,47.5}^{s/(4+s)}n_{\rm ext,3}^{(3-s/2)/(4+s)}\nonumber\\
&\times&\epsilon_{B,-2}^{(1+s/2)/(4+s)}\zeta_{e,-0.5}^{2(s-1)/(4+s)}t_{6}^{(2-3s)/(4+s)}.
\end{eqnarray}
Thus, the SSA process can be relevant in the GHz and lower-frequency bands at early times of $t\lesssim10^6$~s.

\section{Implications for Fast Radio Bursts}
Our paper has considered a general setup which is often used in models for FRBs and luminous SNe (including SLSNe).  
In this section, we focus on some implications that are specific to FRBs.  

In Section~5.1, we discuss implications of the burst-in-bubble model for FRBs and constraints on their 
emission regions. FRBs should be attributed to coherent radio emission, which suggests that the emission radius 
and/or the wind Lorentz factor are large enough.  

Remarkably, for some FRBs with high S/N, polarization and scattering features have been confirmed, which may indicate 
that the sources are in dense regions of the host galaxies, e.g., in star-forming regions
~\citep{Petroff+15,Masui+15,Katz16}.  The magnetar and RNS scenarios seem consistent with this indication. 

Recently, \cite{2016Natur.531..202S} reported a repeating activity of FRB 121102, while a repeatability is not 
observed in other FRBs.  Given the relatively small observed flux and DM, FRB 121102 may belong to a 
less energetic but more frequent subclass of FRBs located at a closer distance.  
In the burst-in-bubble model, the repeating feature could be attributed to intrinsic episodes of bursting activities or 
inhomogeneous interactions with the nebula. However, detailed studies of this are beyond the scope of this work.   

As shown in the above, detectable HEGFs and subsequent afterglows are possible in the magnetar scenario.  
Thus, it is natural to ask whether the non-observation of giant flares associated with FRBs would give 
strong constraints on the model or not.  We examine this issue in Section~5.2.  

While we have focused on scenarios involving a wind bubble, FRBs may be produced by different progenitors 
but with the same mechanism. Indeed, at present, FRBs could be compatible with multiple physical origins. 
For instance, FRB 150418 is claimed to be hosted in an elliptical galaxy, where the star-formation 
activity would be significantly low~\citep{Keane+16}.  
Although its association is largely disputed~\citep[e.g.,][]{2016arXiv160208086Z,Williams&Berger16,2016arXiv160304880A}, 
we briefly discuss some of the general implications for merger scenarios in Section~5.3.

\subsection{On coherent radio emission}
As mentioned above, FRB mechanisms should involve a coherent emission process.  
One general constraint comes from the induced-Compton scattering, although details are sensitive to 
the angle between a radio pulse and the relativistically moving plasma~\citep{2013PTEP.2013l3E01T}.    
The induced-Compton scattering time is longer than the dynamical time of the photon beam when~
\citep{1978MNRAS.185..297W,2008ApJ...682.1443L}
\begin{equation}
\tau_{\rm ind}^{\rm int}\approx\frac{3\sigma_{\rm T} n'_{we} I_{\rm FRB}\Gamma_0 r_0}{2 m_{\rm e}\gamma'_{\rm T}\nu^2}\lesssim1,
\end{equation}
where $n'_{we}$ is the comoving lepton density of the bursting flow and $\gamma'_{\rm T}$ is the temperature of electrons and positrons.  
The radiation intensity in the comoving frame can be connected to the observed isotropic luminosity of FRBs, 
$I_{\rm FRB}\approx L_{\rm FRB}/(64 \pi^2 \Gamma_0 r^2 \nu)$. 
Here we assume that the size of the emission region is $l'_0 \approx\Gamma_0c\delta t\gtrsim r_0/(2\Gamma_0)$. 
The induced-Compton scattering can be neglected if 
\begin{eqnarray}
r_0&\gtrsim& 3.8\times{10}^{16}~{\rm cm}~{\mathcal E}_{48}^{1/3}L_{\rm FRB, 43}^{1/3}{\gamma'}_{\rm T}^{-2/3}\nonumber\\
&\times&{\Gamma}_{0,6}^{-2/3}\nu_9^{-1}{\Delta t}_{-3}^{-1/3}\sigma_0^{-1/3}
\end{eqnarray}
in the magnetar scenario,
\begin{eqnarray}
r_0&\gtrsim& 8.3\times{10}^{14}~{\rm cm}~{\mathcal E}_{43}^{1/3}L_{\rm FRB, 43}^{1/3}{\gamma'}_{\rm T}^{-2/3}\nonumber\\
&\times&{\Gamma}_{0,6}^{-2/3}\nu_9^{-1}{\Delta t}_{-3}^{-1/3}\sigma_0^{-1/3}
\end{eqnarray}
in the RNS scenario, and
\begin{eqnarray}
r_0&\gtrsim& 3.8\times{10}^{15}~{\rm cm}~{\mathcal E}_{45}^{1/3}L_{\rm FRB, 43}^{1/3}{\gamma'}_{\rm T}^{-2/3}\nonumber\\
&\times&{\Gamma}_{0,6}^{-2/3}\nu_9^{-1}{\Delta t}_{-3}^{-1/3}\sigma_0^{-1/3}
\end{eqnarray}
in the MWD scenario. 
Thus, the FRB emission would be produced outside the light cylinder $R_{\rm lc}\equiv c/\Omega=cP/(2\pi)$ 
in all the three scenarios, unless the magnetization parameter is so large and the plasma density is small enough. 

While the mechanism of the coherent emission is highly uncertain, one of the interesting possibilities is 
the synchrotron maser mechanism~\citep{Lyubarsky14}.  In this case, the characteristic frequency of the 
radio emission is estimated as $\nu_{\rm maser}\approx(eB_*R_*)/(2^{3/2}\pi m_ec\Gamma r_0)$. 
If $r_0\approx R_{\rm nb}$ is assumed, we have
\begin{eqnarray}
{\nu}_{\rm maser}&\simeq&6.0\times10^{7}~{\rm Hz}~{\mathcal E}_{\rm 48}^{-1/4}P_{i,-0.5}^{2/5}B_{*,15}^{1/2}\nonumber\\
&\times&M_{\rm ext,0.5}^{1/5}V_{\rm ext,8.75}^{-3/5}T_{\rm 10yr}^{-3/2}{\Delta t}_{-3}^{1/4}
\end{eqnarray}
in the magnetar scenario,
\begin{eqnarray}
{\nu}_{\rm maser}&\simeq&4.4\times10^{6}~{\rm Hz}~{\mathcal E}_{\rm 43}^{-1/4}P_{i,-3}B_{*,12.5}^{1/2}\nonumber\\
&\times&M_{\rm ext,0.5}^{1/2}T_{\rm 10yr}^{-3/2}{\Delta t}_{-3}^{1/4}
\end{eqnarray}
in the RNS scenario, and
\begin{eqnarray}
{\nu}_{\rm maser}&\simeq&2.9\times{10}^4~{\rm Hz}~{\mathcal E}_{\rm 45}^{-1/4}P_{i,1}^{-1/5}B_{*,9}^{11/10}\nonumber\\
&\times&M_{\rm ext,-3}^{1/5}V_{\rm ext,9}^{-3/5}T_{\rm 10yr}^{-6/5}{\Delta t}_{-3}^{1/4}
\end{eqnarray}
in the MWD scenario.  As proposed by \citet{Lyubarsky14}, this mechanism can be promising in the magnetar scenario.  
In the other two cases, the typical frequency would be smaller than the observed one, unless $r_0\ll R_{\rm nb}$.

\subsection{Possible constraints on the magnetar scenario}
\begin{table*}\label{tab:list}
\caption{Constraints on soft gamma-ray counterparts of FRBs. In the middle, the DM in a host galaxy 
(including the source environment) is estimated assuming the source distance.  In the right, the source 
distance is estimated assuming the typical DM in elliptical galaxies.}
\begin{tabular}{c*{3}{c} | c*{1}{c} | c*{2}{c}}
\hline
\multirow{2}{*}{Name}
 & $\rm DM_{obs} \ {}^{\rm a} $ & $\int d \varepsilon \, \varepsilon\phi_{\gamma}^{\rm max} \ {}^{\rm b} $ & $\rm DM_{MW} \ {}^{\rm c} $ & $\rm DM_{host, 300Mpc} \ {}^{\rm d} $ & ${\mathcal E}_{\gamma,300{\rm Mpc}}^{\rm max}  \ {}^{\rm e} $ & $z_{\rm ell}  \ {}^{\rm f}$  &  ${\mathcal E}_{\gamma, \rm ell}^{\rm max}  \ {}^{\rm g} $    \\ 
 & $\rm [pc \ cm^{-3}]$ & $\rm [erg~cm^{-2}]$  & $\rm [pc \ cm^{-3}]$ & $\rm [pc \ cm^{-3}]$ & [erg] & & [erg] \\
\hline
\ FRB 010724 & 375 & $2.0\times 10^{-7}$ & 45  & 258 & $2.1 \times 10^{48}$  & 0.29 & $5.7 \times 10^{49} $& \\
\ FRB 110220 & 944 & $2.0\times 10^{-7}$ & 39 & 870 & $2.1 \times 10^{48}$ & 0.87 & $7.9  \times 10^{50}$ & \\
\ FRB 130729 & 861 & $2.0\times 10^{-8}$ & 31 & 790 & $2.1 \times 10^{47}$ & 0.80 & $6.3 \times 10^{49}$ & \\
\ FRB 011025 & 790 & $2.0\times 10^{-7}$ & 570 & 141 & $2.1 \times 10^{48}$ & 0.17 & $1.8 \times 10^{49}$ & \\
\ FRB 131104 & 779 & $1.0\times 10^{-8}$ & 190 & 533 & $1.1 \times 10^{47}$ & 0.55 & $1.3 \times 10^{49}$ & \\
\ FRB 121002 & 1629 & $1.0\times 10^{-8}$ & 260 & 1363 & $1.1 \times 10^{47}$ & 1.3 & $1.2 \times 10^{50}$ & \\
\ FRB 090625 & 900 & $1.0\times 10^{-8}$ & 39 & 823 & $1.1 \times 10^{47}$ & 0.83 & $3.8 \times 10^{49}$ & \\
\ FRB 110703 & 1104 & $1.0\times 10^{-8}$ & 33 & 1046 & $1.1 \times 10^{47}$ & 1.0 & $6.1 \times 10^{49}$ & \\
\ FRB 130626 & 952 & $2.0\times 10^{-8}$ & 54 & 862 & $2.1 \times 10^{47}$ & 0.86 & $7.7 \times 10^{49}$ & \\
\ FRB 140514 & 563 & $2.0\times 10^{-8}$ & 37 & 466 & $2.1 \times 10^{47}$ & 0.49 & $2.0 \times 10^{49}$ & \\
\ FRB  130628 & 470  & $1.0\times 10^{-8}$ & 53 & 350 & $1.1 \times 10^{47}$ & 0.38 & $5.3 \times 10^{48}$ & \\
\ FRB 121102 & 557 & $2.0\times 10^{-8}$ & 190 & 297 & $2.1 \times 10^{47}$ & 0.33 & $7.6 \times 10^{48}$ & \\
\ FRB 110626 & 723 & $2.0\times 10^{-7}$ & 56 & 616 & $2.1 \times 10^{48}$ & 0.63 & $3.6 \times 10^{50}$ & \\
\ FRB 120127 & 553 & $2.0\times 10^{-8}$ & 33 & 460 & $2.1 \times 10^{47}$ & 0.48 & $1.9 \times 10^{49}$ & \\
\hline
\end{tabular}

\begin{flushleft}
{\small ${}^{\rm a, b}$ Observed dispersion measures and upper limit on the soft gamma-ray fluence~\citep[][and references therein]{Tendulkar+16}. 
${}^{\rm c}$ Calculated dispersion measures in our Galaxy based on the NE2001 model~\citep{NE2001a,NE2001b}. 
${}^{\rm d,e}$ Calculated dispersion measures in the host galaxy and upper limit on the intrinsic soft gamma-ray radiation energy 
with the assumption that the distance is $300 \ \rm Mpc$ from the Earth. 
${}^{\rm f,g}$ Calculated source redshifts and upper limits on the intrinsic soft gamma-ray radiation energy with the assumption that the host galaxy is elliptical. 
}
\end{flushleft}

\label{tab:summary}
\end{table*}

So far there are no $\gamma$-ray counterpart detections associated with FRBs.  \cite{Tendulkar+16} summarized 
the fluence upper limits obtained by Konus-Wind~\citep{Aptekar+95}, the Burst Alert Telescope (BAT) onboard 
{\it Swift}~\citep{Barthelmy+05}, and the GRB Burst Monitor (GBM) onboard {\it Fermi}~\citep{Meegan+09}. 
Assuming that these results are correct, we here convert them into upper limits of the intrinsic isotropic energy of soft gamma-rays 
by setting source distance in two different ways (see Table \ref{tab:summary}).

First, we simply assume a source distance of $300 \ \rm Mpc$.  In general, an observed DM can be divided into 
\begin{equation}\label{eq:DM}
{\rm DM}_{\rm obs} =  {\rm DM}_{\rm MW} + {\rm DM}_{\rm halo} +  {\rm DM}_{\rm IGM} + \frac{{\rm DM}_{\rm host}}{1+z},
\end{equation}
where the first, second, and third term corresponds to contributions from our Galaxy, the intergalactic medium, 
and the host galaxy including regions close to the source, e.g., the nebula and SN ejecta, respectively. 
We use the NE2011 model for calculating ${\rm DM}_{\rm MW}$~\citep{NE2001a,NE2001b} and set ${\rm DM}_{\rm halo} 
= 30 \ \rm pc \ cm^{-3}$.  The intergalactic contribution is estimated from
\begin{equation}
{\rm DM_{IGM}} = \frac{3c H_0 \Omega_{\rm IGM}}{8 \pi G m_{\rm p}} \int^{z}_{0} \frac{(1+z')f_e(z')dz'}{[(1+z)^3 \Omega_{\rm m} + \Omega_\Lambda]^{1/2}},
\end{equation}
with the cosmological parameters $\Omega_{\rm m} = 0.3089$, $\Omega_\Lambda = 0.6911$, $\Omega_{\rm b} = 0.0486$, 
and $H_0 = 67.74 \ \rm km \ s^{-1} \ Mpc^{-1}$~\citep{Planck2015}.
We also assume $f_e = 0.88$ and $\Omega_{\rm IGM} = 0.9 \times \Omega_{\rm b}$. 
At $300 \ \rm Mpc$ ($z = 0.0644$), ${\rm DM}_{\rm IGM, 300Mpc} = 65 \ \rm pc \ cm^{-3}$, 
and a significant fraction of the observed DM needs to originate from the host galaxy including the source 
environment such as the nebula and baryonic ejecta. 

The estimated upper limits of the intrinsic isotropic energy of soft gamma-rays ranges from 
${\mathcal E}_{\gamma, 300{\rm Mpc}}^{\rm max} \sim 10^{47}-{10}^{48} \ \rm erg$, 
which is larger than the energy of the giant flare from SGR 1806-20 but close to values expected in the 
hyper-flare scenario. 
Thus, if soft gamma-ray emission is produced concurrently with magnetic reconnections~
\citep{1995MNRAS.275..255T,2001ApJ...561..980T, 2010MNRAS.407.1926G,2016MNRAS.456.3282E}, 
the absence of correlations between FRBs and hyper-flares can constrain the magnetar scenario (if 
the soft gamma-ray emission is ubiquitous). 
Based on these results, we encourage further soft gamma-ray searches especially with {\it Swift}. 
Possible detection of FRB counterparts with gamma-ray energies exceeding values of Eq.~(33) can exclude
magnetar and MWD models, in which the magnetic energy of remnants such as NSs and WDs is used.  
Non-detection is also useful. 
If the nebula and ejecta contribute more significantly to the DM, the limits would become tighter, 
so that the hyper-flare model is disfavored.  
This is because the distance to FRBs would need to be smaller than $\sim300$~Mpc in this case.  
Of course, an association with a dimmer class of magnetically-induced bursts is not ruled out, 
even for an FRB distance of $d\lesssim 300 \ \rm Mpc$.

\subsection{Possible constraints on merger scenarios}
Our burst-in-bubble model is applicable not only to models involving a SN explosion but also to some 
models involving compact mergers, as considered in the MWD scenario. 
Even in merger scenarios, mass ejection is expected around the coalescence time.  
The total ejected mass is rather uncertain and depends on the binary parameters; for example, $M_{\rm ext} 
\sim 10^{-5}-10^{-2} \ M_\odot$ for a NS-NS binary based on numerical simulations
~\citep[see, e.g.,][]{2014PhRvD..90d1502K,2015PhRvD..91f4059S}.  
The ejecta velocity is expected to be $V_{\rm ext}\approx0.3 c$ for a NS-NS merger and 
$V_{\rm ext}\approx0.06 c$ for a WD-WD merger. 
However, as long as we consider FRB emission at early times with $T\lesssim1$~yr, the corresponding electron 
density is high enough to prevent FRB emission from escaping the system. Thus, FRBs need to be produced 
as precursors before the actual merger~\citep[e.g.,][]{2013PASJ...65L..12T,2013ApJ...768...63L}.  
Alternatively, since  the  mass ejection in general occurs in a highly anisotropic manner, FRBs may be
possible when the magnetic dissipation occurs outside the ejecta, if relativistic Poynting-dominated outflows 
get ahead of the slower ejecta~\citep[e.g.,][]{Kashiyama+13}.


In such situations, contributions of the source environment to ${\rm DM}_{\rm host}$ are expected to be small. 
Assuming that ${\rm DM}_{\rm host}$ is dominated by the host galaxy contribution, in Table~1, 
we use $ {\rm DM}_{\rm host} = 30 \ \rm pc \ cm^{-3}$ that is typical for elliptical galaxies.   
Then, the redshifts of the host galaxies are estimated from Eq. (\ref{eq:DM}) as $z_{\rm ell} \sim 0.2-1$. 
We finally derive upper limits on the intrinsic isotropic energy of the soft $\gamma$-rays ranging from 
${\mathcal E}_{\gamma,\rm ell}^{\rm max} \sim 5 \times 10^{48} \ \rm erg$ to ${\mathcal E}_{\gamma,\rm ell}^{\rm max} \sim 5 \times 10^{50} \ \rm erg$. 
This would exclude an association of an FRB with on-axis short GRBs, at least for several events.  

In merger scenarios, the mass ejection leads to a broadband afterglow emission that may be detectable with 
dedicated follow-up observations~\citep{2014PASJ...66L...9N}~\citep[and see also][for the discussion on 
possible afterglows]{2014ApJ...792L..21Y}, 
which has also been of interest for giving rise to counterparts of gravitational waves.
The deceleration time of the merger ejecta is estimated to be $T_{\rm ST}\simeq 2.1~{\rm yr}~
M_{\rm ext, -3}^{1/3} n_{\rm ism}^{-1/3} V_{\rm ext, 10}^{-1}$.  
In the case of the dynamical ejecta of a NS-NS merger, the peak flux is in the range $\sim 10^{26-28} \ \rm erg 
\ s^{-1} \ Hz^{-1}$ in the $\sim 0.1-1 \ \rm GHz$ bands~\citep[e.g.,][and references therein]{hotoke15}, 
and the emission is slower and dimmer for WD-WD merger with a slower ejecta velocity.  
The radio emission is stronger if a long-lived magnetar exists~\citep{2014MNRAS.437.1821M}, which may already 
be ruled out for the two kilonova/macronova candidates~\citep{2016ApJ...819L..22H}.  
More rapid radio afterglows fading within $100$ days can be also produced if a short GRB jet is launched 
after the NS-NS merger.  In any of the references mentioned above, the radio signature is qualitatively different 
from the one considered in the present work.

Finally, we note that merger scenarios discussed above cannot explain a repeating FRB such as that 
reported by \citet{2016Natur.531..202S}.  However, as argued by many authors, 
FRBs may have multiple origins and these models could still be viable to explain the other FRBs. 

\section{Summary}
We have studied the various roles played by a tenuous wind bubble formed by relativistic winds from a rotating 
compact remnant such as a NS or WD.  Such a setup is commonly expected for young NSs and WDs after stellar 
core-collapses or mergers, respectively, and it is often employed in models for FRBs as well as SLSNe and GRBs. 
Throughout this paper, we considered three examples, (a) the magnetar scenario, (b) the RNS scenario, and 
(c) the MWD scenario.  Our results are summarized as follows.

(i) First, we calculated the quasi-steady nebular emission of extragalactic pulsar wind nebulae, based on 
parameters of Galactic pulsar wind nebulae. 
The nebula's emission itself is difficult to detect in the magnetar and MWD scenarios.  
On the other hand, in the RNS scenario, the GHz radio flux at $T\sim100$~yr can be $\sim1$~mJy for an event 
at $d\sim100$~Mpc, and the related emission is detectable with present radio telescopes such as VLA, SMA, and ALMA. 
In particular, higher-frequency emission in the sub-mm band would provide a way to test pulsar-driven SN model and FRBs. 
Then, we studied the effects of the nebula's emission on the attenuation of high-energy gamma-rays and radio waves.  
We showed that the escape of high-energy gamma-rays, including TeV photons, is possible in all three scenarios.  
The radio emission can be severely absorbed in the nebula and in the external ejecta, but the system can be 
transparent to GHz radio waves at time $T\gtrsim10-100$~yr.  Thus, if FRBs originate from magnetars or RNSs, 
their age is expected to be $T\gtrsim10-100$~yr.   

(ii) Second, we suggested HEGFs caused by possible magnetic dissipation in the nebula.  
An impulsive relativistic outflow, which propagates in the wind and is quickly decelerated by interactions with the nebula, 
can boost a population of pre-existing non-thermal particles. 
The duration HEGFs may be as short as milliseconds but can also be longer, depending on the expansion of the outflow.
For the magnetar scenario, HEGFs from the forward shock may occur in the fast-cooling regime, and TeV 
gamma-rays from nearby events with $\lesssim10-100$~Mpc may be detected with HAWC.  
CTA could detect more distant HEGFs up to $\sim0.1-1$~Gpc although its field-of-view is much smaller. 
On the other hand, only a fraction of the energy would be converted into gamma-rays in the RNS and MWD 
scenarios, so detecting HEGFs from these would be more challenging.   We also considered possible flaring 
emission from the reverse shock, which may lead to additional gamma-ray emission.  

(iii) Third, we considered the subsequent afterglow emission following HEGFs and possible FRBs.  
In the magnetar scenario, the associated radio emission from the external forward shock is detectable with 
dedicated follow-up observations, for a nearby event at $\sim10-100$~Mpc. 

(iv) While our setup is not sensitive to details of the FRB mechanism, we discussed specific implications 
for FRBs.  The induced-Compton scattering implies that the emission regions are far beyond the 
light cylinder for the  burst-in-bubble models considered here. 
We also considered limits on hyper-flares expected in the magnetar scenario.  
The limits are consistent with the model, as long as they do not exceed ${\mathcal E}_{\rm mag}\sim{10}^{48}$~erg.
But these limits should be tighter for younger nebulae since electrons and positrons in the nebula and/or ejecta can 
significantly contribute to the DM.   

\section*{Acknowledgments}
K.~M. thanks Hiroya Yamaguchi for useful discussion. 
K.~K. is supported by NASA through Einstein Postdoctoral Fellowship grant number PF4-150123 awarded by the 
Chandra X-ray Center, operated by the Smithsonian Astrophysical Observatory for NASA under contract NAS8-03060.
P.~M. acknowledges partial support by NASA NNX13AH50G.

\appendix
\section[]{On the detection of quasi-steady nebula emission}
In this Appendix, we briefly discuss radio signatures of the quasi-steady nebular emission described in Section~2.2. 
For the calculations, we use the code developed by \citet{2015ApJ...805...82M}, taking also into account the 
effect of residual electron-positron pairs, which are relevant for the emission after the spin-down time. 
Our results on the light curves are shown in Figs.~11-13. 

\begin{figure*}
\includegraphics[width=0.32\linewidth]{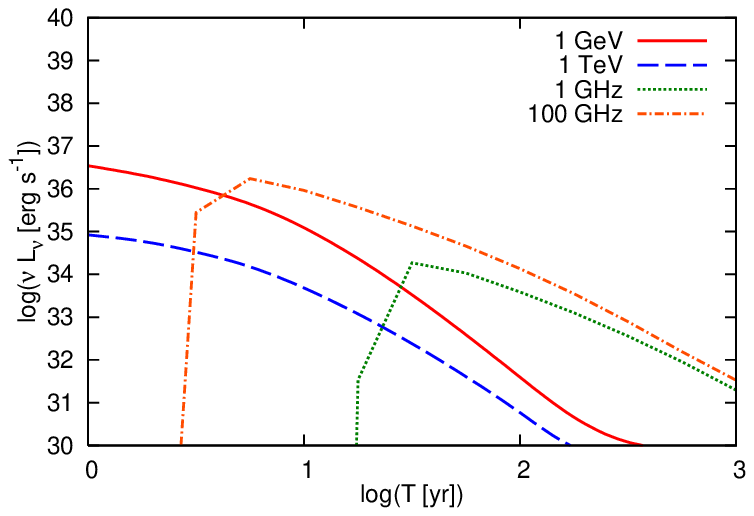}\hfill
\includegraphics[width=0.32\linewidth]{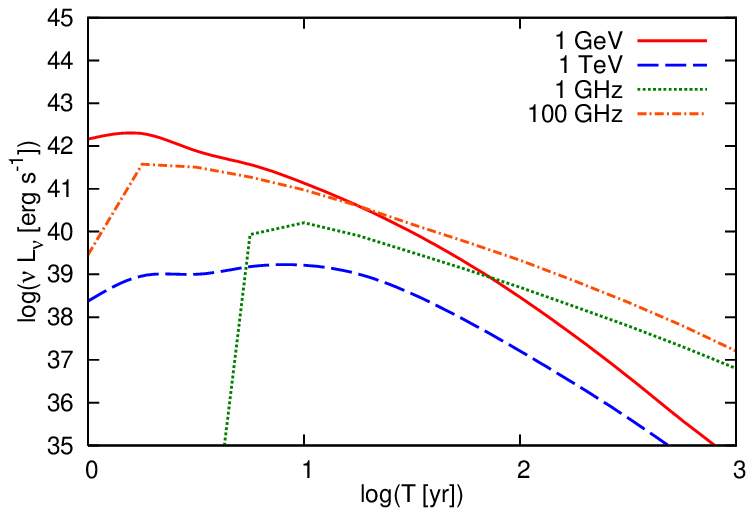}\hfill
\includegraphics[width=0.32\linewidth]{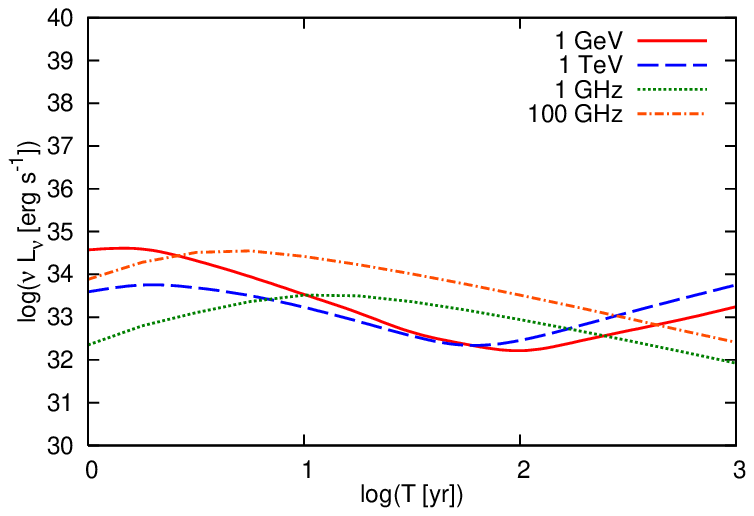}
\caption{Corresponding to Fig.~2, light curves of radio and gamma-ray emission from a very young nebula are shown 
for the magnetar scenario (left panel), RNS scenario (middle panel), and MWD scenario (right panel). 
}
\end{figure*}

In the magnetar scenario, the nebular synchrotron emission is too weak to be detected at late times, 
but embryonic emission (at $T\lesssim{\rm a~few}$~yr) can be of interest at high frequencies. 
However, as shown in Section~2.3, both the radio and sub-mm emission will be masked by the free-free 
absorption for our parameter set, so only X-rays and high-energy (GeV-TeV) gamma-rays would be 
detectable~\citep{2015ApJ...805...82M}.  On the other hand, for our parameters of the RNS scenario, 
the GHz radio flux around $T\sim100$~yr is $F_{\nu}^{\rm nb}\sim1~ {\rm mJy}~d_{26.5}^{-2}$ 
(see the middle panel of Figs.~2 and 12), so RNS at $\lesssim100$~Mpc nebulae would be detectable 
with current radio telescopes. 

Our results shown in Figs.~2 and 12 imply that searching for radio counterparts of pulsar-driven SNe, 
including SLSNe, is relevant especially at higher frequencies, where the absorption is less relevant.  
The Atacama Large Millimeter/sub-millimeter Array (ALMA) may detect the nebular emission from the RNS. 
Although the detectability is sensitive to $P$ (that increases at $T>T_{\rm sd}$), for our parameter set 
of the RNS scenario, the 100~GHz radio flux around $T\sim10$~yr reaches $F_{\nu}^{\rm nb}\sim0.8~{\rm mJy}~d_{27.5}^{-2}$, 
which exceeds the ALMA sensitivity of $\sim0.01$~mJy as well as that of the SMA array. 

The pulsar-driven SN model requires that almost all the spin-down energy is dissipated, 
and a significant fraction of that energy needs to be converted into the observed SN emission. 
Thus, the  radio and sub-mm signals can be used as relevant probes.  
The rate of SLSNe is only $\sim10~{\rm Gpc}^{-3}~{\rm yr}^{-1}$~\citep{Gal-Yam_2012}. Thus, the number 
of their radio counterparts within 100 Mpc with $T\lesssim10-100$~yr is about $0.3-3$, 
so non-detections with radio surveys are not constraining.  
However, future dedicated follow-up surveys should provide a powerful test.  This is especially the case 
for sub-mm telescopes, which have a small field of view. 

The quasi-steady nebular emission can be more relevant for testing models for FRBs. The FRB rate is 
$\sim{10}^{4}~{\rm Gpc}^{-3}~{\rm yr}^{-1}$, leading to the number of their radio counterparts 
within 100 Mpc with $T\lesssim10-100$~yr being about $300-3000$.  This is especially the case for a 
RNS model such that the FRBs are powered by the spin-down luminosity rather than the magnetic energy.  
Note that the rotation period increases after the spin-down time $T_{\rm sd}$, and the spin-down luminosity 
for our RNS parameters is $L_{\rm sd}\sim{10}^{43}~{\rm erg}~{\rm s}^{-1}$ at $T\sim10$~yr, which is 
comparable to the observed FRB luminosity.  
We expect that the RNS model for FRBs is strongly constrained for RNSs with $P_i\sim1-10$~ms, because of 
the absence of such luminous transients in the radio sky 
(as long as a significant fraction of the spin-down energy is dissipated around the nebula).  
Note that such nebular emission should be non-thermal, which can be discriminated from Galactic 
thermal sources. Discriminating them from active galactic nuclei would be more difficult, and it 
would need information on counterparts at other wavelengths and/or variability time scales.

\section[]{Comments on FRB 121102}

Recently, a radio counterpart of FRB 121102 was reported via the observations by the VLA and the European VLBI Network (EVN), respectively~\citep{2017Natur.541...58C,2017ApJ...834L...8M}. \cite{2017ApJ...834L...7T} identified their host galaxy with redshift $z\approx0.19$. For the first time, they have established an FRB as a cosmological event. They also found a quasi-steady radio source with an observed flux of $\sim0.2$~mJy at $\sim1-10$~GHz, which corresponds to a luminosity of $\sim{10}^{39}~{\rm erg}~{\rm s}^{-1}$. Remarkably, this is consistent with our model, as seen in the middle panel of our Figure~\ref{fig2} for the RNS scenario. Indeed, the parameters, $B_{*}=10^{12.5}$~G and $P_i=1$~ms, lie in the allowed parameter space shown in \cite{2017arXiv170104815K}. Note that SSA is the most important process that suppresses the synchrotron flux at sufficiently low frequencies. Also, in general, the energy source of FRB emission itself does not have to be rotation energy that is responsible for the nebular emission~\citep[e.g.,][]{2014A&A...562A.137F}.

While other possibilities such as emission from low-luminosity active galactic nuclei are not ruled out, this discovery encourages our proposal of follow-up observations of pulsar-driven SN candidates, especially at submm frequencies. SLSNe are of particular interest, where $\sim1-10$~yr time scale observations at high frequencies are crucial. With the MHD-motivated spin-down formula, \cite{Kashiyama+16} showed that SLSNe favor $B_*\sim{10}^{13}-{10}^{14}$~G and $P_i\sim{\rm a~few}$~ms~\citep[see Figure~6 of][]{Kashiyama+16}. The parameters favored by FRB 121102 are not too far from them, so searches for radio counterparts will give us important information on the relationship between FRBs and pulsar-driven SNe, including SLSNe. Also, as suggested by \cite{Kashiyama+16} (see their Figure~9), ``hard'' X-rays should also serve as promising signals for the parameters that may account for SLSNe and FRBs. 

\bibliography{ref}

\bsp

\label{lastpage}

\end{document}